\documentclass[10pt, a4paper, twocolumn, aps, pra, showpacs, longbibliography, nofootinbib,superscriptaddress]{revtex4-2}

\usepackage{ucs}
\usepackage{amsmath}
\usepackage{amsfonts}
\usepackage{amssymb}
\usepackage{mathtools}
\usepackage{makeidx}
\usepackage{cellspace,booktabs}
\usepackage{natbib}
\usepackage{lipsum}
\usepackage{bm}
\usepackage{bbm}
\usepackage{relsize}
\usepackage{hyperref}
\hypersetup{colorlinks = true, linkcolor=magenta,citecolor=blue, urlcolor=magenta, bookmarksnumbered =  true}
\usepackage[usenames,dvipsnames]{color}
\definecolor{light-gray}{gray}{0.55}

\usepackage{microtype}

\usepackage{graphicx}

\newcommand{\ssm}{\rm\scriptscriptstyle}

\newcommand{\parspace}{\vspace{0.4cm}}

\begin{document}
%TC:ignore
\begin{abstract}
The two-dimensional electron gas is 
of fundamental importance in quantum many-body physics.
We study a minimal extension of this model with $C_4$ (as opposed to full rotational) symmetry and an 
electronic dispersion 
with two valleys with anisotropic effective masses. 
Using variational Monte Carlo simulations, we find a broad intermediate range of densities 
with a   metallic valley-polarized, spin-unpolarized  
ground state.  Our results are of direct relevance to the recently discovered ``nematic'' state in AlAs quantum wells. For the effective mass anisotropy relevant to this system, $m_x/m_y\approx 5.2$, we obtain a transition from an anisotropic metal to a valley-polarized metal at $r_s \approx 12$ (where $r_s$ is the dimensionless Wigner-Seitz radius). At still lower densities, we find a (possibly metastable) 
valley and spin-polarized state with a reduced electronic anisotropy.
\end{abstract}

\date{\today}
\author{Agnes Valenti}
\affiliation{Institute for Theoretical Physics, ETH Zurich, CH-8093, Switzerland}
\author{Vladimir Calvera}
\affiliation{Department of Physics, Stanford University, Stanford, CA 94305, USA}
\author{Steven A. Kivelson}
\affiliation{Department of Physics, Stanford University, Stanford, CA 94305, USA}
\author{Erez Berg}
\affiliation{Department of Condensed Matter Physics, Weizmann Institute of Science, Rehovot 76001, Israel}
\author{Sebastian D. Huber}
\affiliation{Institute for Theoretical Physics, ETH Zurich, CH-8093, Switzerland}

\title{Nematic metal in a multi-valley electron gas: \\ Variational Monte Carlo analysis and application to AlAs}

\maketitle
%TC:endignore

The two-dimensional electron gas (2DEG) serves as a starting point for the description of a plethora of intriguing quantum phases \cite{ando1982electronic}. 
Among other things, it exhibits a transition from a fluid phase (or phases) at high density, $n$, to an insulating Wigner crystal phase at low $n$ where the Coulomb repulsion dominates over the kinetic energy.  In order to  describe this transition, methods which faithfully capture fluctuations beyond mean-field approaches are required. Quantum Monte Carlo methods such as variational and diffusion Monte Carlo have emerged as highly successful in the treatment of the 2DEG \cite{tanatar1989ground, kwon1993effects, rapisarda1996diffusion, attaccalite2002correlation, gori2003two, drummond2009phase, marchi2009correlation}. Unfortunately, the simplicity of the uniform isotropic electron gas is also reflected in its phase diagram: While an ongoing debate does not preclude the existence of subtle forms of order \cite{spivak2004phases, jamei2005universal, raghu2011superconductivity, kim2022interstitial}, there is no conclusive evidence for a {\em metallic, symmetry-broken state} stabilized by interactions according to quantum Monte Carlo calculations so far \cite{ drummond2009phase, loos2016uniform, marchi2009correlation}. 
In the present manuscript we show how a minimal extension of the two-dimensional electron gas can lead to an intermediate symmetry-broken {\em metallic} state.

With the advent of two-dimensional van der Waals materials~\cite{zhou2021superconductivity, seiler2022quantum, cao2018unconventional, chen2019signatures, park2021tunable, bistritzer2011moire, cao2018correlated, shen2020correlated, zhao2023gate, mak2022semiconductor}, the importance of multi-valley physics including several anisotropic Fermi-pockets has %become apparent.
come into renewed focus.  However, in most of these systems, such as graphene- or transition-metal-dichalcogenide multilayers, the complexity of the observed physics poses a challenge to accurate theoretical modelling. Recent QMC studies \cite{hofmann2022fermionic, Yang2023Metal} on 2D van der Waals materials showcase the potential of accurate treatments of the interactions in strongly correlated systems. Here, we aim at an in-depth study of a simpler model, which however shares a key feature with the aforementioned layered materials: the existence of a valley degree of freedom. Moreover, to account for more intricate effects arising from a non-trivial Fermi surface, we consider a model with two {\em anisotropic} pockets, related by $90^\circ$ rotation. 

Beyond its theoretical appeal, 
our model provides a good description of the 2DEG in ultra clean AlAs quantum wells~\cite{shayegan2006two}. 
 Recent experiments~\cite{hossain2020observation, hossain2021spontaneous, hossain2022anisotropic}  
have reported the existence of at least three distinct phases that appear in this system upon decreasing $n$: i) The symmetric Fermi liquid gives way to a {\em valley-polarized} phase appearing around the Wigner-Seitz radius $r_s\propto 1/\sqrt{n}\approx 20$ \cite{footnote1, hossain2021spontaneous}; ii) an insulating phase which is spin unpolarized is stabilized for $27 < r_s < 35$ and iii) An {\em insulating spin- and valley-polarized state} appears at $r_s \sim 35$ \cite{hossain2020observation, hossain2022anisotropic}.

As we shall elaborate below, Hartree-Fock mean-field theory cannot explain these findings, even at a qualitative level. 
The experimental observations are further in apparent disagreement \cite{Ahn2022} with the quantum Monte Carlo results of the 2D {\em isotropic} electron gas that do not predict any polarized fluid phase, even in its multi-valley variant \cite{marchi2009correlation}. On the contrary, introducing multi-valley components alone has the rather opposite effect of further stabilizing the paramagnetic fluid.
These observations call for a quantum Monte Carlo study taking into account both the multi-valley nature of the system {\em and} the anisotropy of the Fermi surface.

We study the 2D multi-valley anisotropic electron gas beyond mean-field by performing variational Monte Carlo (VMC) calculations using a Slater-Jastrow-Backflow trial wave-function. We focus on densities larger than the expected transition to a Wigner crystal and therefore explore different fluid phases. We find that valley-polarization emerges as a result of both multi-valley physics and anisotropy and thereby resolve the apparent contrast to the isotropic 2D electron gas.
For the special case of AlAs, we are able to approximately match the relevant density regime at which valley-polarization is experimentally observed.
We  further predict 
a correlation-induced renormalization of the Fermi surface, which results in a reduced anisotropy relative to the band value. 

%\parspace

 %model

\begin{figure}[b]
\includegraphics[scale=1]{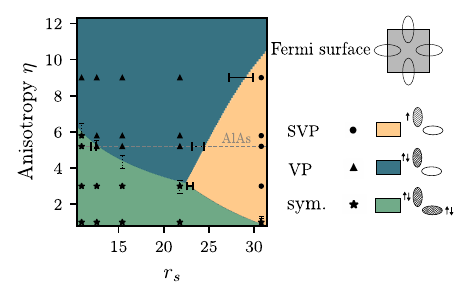}
\caption{
The phase diagram of a 2D multi-valley anisotropic electron gas as a function of $r_s$ and the anisotropy $\eta$. The simulations were performed on a $r_s-\eta$ grid, indicated by markers. The shape of the markers denote the polarization of the ground state. The phase boundaries resulting from interpolation in $r_s$ ($\eta$) of the discrete simulated points are marked with a horizontal (vertical) error bar, denoting the uncertainty in the energy crossing. The error bars result from uncertainty in the finite-size scaling as well as $r_s-\eta$ interpolation.}
\label{fig:fig1}
\end{figure}

{\it Multi-valley anisotropic electron gas.}
We consider a system with two conduction band minima that lie at the $X$-points of the Brillouin zone. For small electron densities, the conduction band electrons near the minima can be treated as two separate flavours, or isospins, of a 2D electron gas. We include anisotropy by allowing for an elliptic shape of the valleys, with distinct longitudinal and transverse effective masses. Concretely, we consider the Hamiltonian
\begin{align}
\label{eq:H}
H=-\sum \limits_{i} \frac{1}{2m^{*}}\big(\eta^{\tau_i/2} \partial_{i,x}^2+ \eta^{-\tau_i/2} \partial_{i,y}^2 \big) + \sum \limits_{i<j} V(|{\bf r}_i- {\bf r}_j|),
\end{align}
where $i,j$ run over all electrons, $\tau_i \in \{+1, -1 \}$  denotes the valley flavor (i.e. the isospin) of electron $i$ \cite{SI}, and the effective mass is given by $m^{*}$. The parameter $\eta$ controls the anisotropy of the system: $\eta\neq 1$ corresponds to an elliptical shape of the Fermi surface, as schematically depicted in Fig.~\ref{fig:fig1}. 
%Do we want to explain/justify the choice?
For computational simplicity, we consider a Coulomb interaction that is screened by two symmetric metal gates 
\begin{align}
V(|{\bf r}_i-{\bf r}_j|)&=\frac{1}{(2\pi)^2}\int {\rm d} {\bf q}\, {\rm e}^{i{\bf q} \cdot ({\bf r}_i-{\bf r}_j)}v({\bf q}), \\
v({\bf q})&= \frac{e^2}{2 \epsilon_0 \epsilon} \frac{\tanh (d |{\bf q}|)}{|{\bf q}|}.
\label{eq:V}
\end{align}
Here, the distance between the gates is $2d$ and $\epsilon$ %corresponds to the dielectric screening 
is the dielectric constant of the system. The above dual-gate screened interaction decreases exponentially in real-space for long distances as $V(r)\sim{\rm e}^{-r/2d}$ \cite{throckmorton2012fermions}.

{\it Variational Monte Carlo.} %method
We capture correlations beyond mean-field theory by performing VMC calcuations utilizing a trial wave function of Slater-Jastrow-Backflow \cite{jastrow1955many, lee1981green, kwon1993effects, kwon1998effects} type
\begin{align}
\Psi({\bf r}_1, ...,{\bf r}_N)= e^{-J({\bf r}_1, ...{\bf r}_N)}\Psi_D ({\bf r}_1, ...,{\bf r}_N). \label{eq:J_PsiD}
\end{align}
The trial wave-function is separated into the Jastrow pre-factor $e^{-J}$ and an antisymmetric determinantal part $\Psi_D$. Both parts contain parameters that are optimized using stochastic reconfiguration \cite{sorella1998green, sorella2007weak} to minimize 
\begin{align}
    E_{\ssm var}=\frac{\langle \Psi | H | \Psi \rangle}{\langle \Psi |\Psi \rangle}.
\end{align}
Expectation values of interest can then be computed using Monte Carlo sampling of a finite number of electron positions ${\bf r}_i$.

The determinantal part $\Psi_D$ is of the form
\begin{align}
    \Psi_D ({\bf r}_1, ...,{\bf r}_N)&=\prod \limits_{\alpha=\{\tau, \sigma\}} {\rm Det} \big( {\bf M}^{\alpha} \big), \\
    \big({\bf M}^{\alpha}\big)_{ij}&=\varphi_j({\bf r}_i)\propto e^{i {\bf k}_j \cdot {\bf r}_i}.
\end{align}
Given that we consider fluid phases, where translation symmetry fixes the form of the orbitals to be plane waves, the variational degrees of freedom reduce to the choice of which orbitals are filled. 
We choose the Slater determinant to be a filled elliptical Fermi sea \cite{ahn2021anisotropic} whose ratio between semi-axes $\tilde{\eta}_{\ssm{FS}}$ is treated as a variational parameter. We also allow for a symmetry-breaking imbalance of the filling of the isospin  and spin %flavors 
$\tau$ and $\sigma$.

We include correlation effects into our ansatz in two ways. First, we choose a Jastrow factor \cite{jastrow1955many} $J({\bf r}_1,\dots,{\bf r}_N)$ which includes two-body correlations. The Jastrow factor is real and positive and its role is thus to refine the many-body wave-function by effectively keeping electrons apart through the creation of a {\em correlation hole}. We capitalize on the anisotropy of the system in the paramterization of $J({\bf r}_i-{\bf r}_j)$ \cite{whitehead2016jastrow, kim2018qmcpack}, as detailed in \cite{SI}.

A second improvement over a Hartree-Fock type wave-function is achieved through the introduction of a backflow transformation \cite{lee1981green, kwon1993effects, kwon1998effects}: Instead of evaluating the determinantal part of the wave-function $\Psi_D$ at the bare electron coordinates ${\bf r}_i$, the determinant is evaluated at
\begin{equation}
    %{\bf r}_i \to 
    \tilde{\bf r}_i={\bf r}_i+\sum_{j}{\boldsymbol \xi}_{\tau_i,\tau_j}({\bf r}_i-{\bf r}_j).
\end{equation}
The functional form of ${\boldsymbol \xi}_{\tau_i,\tau_j}({\bf r}_i-{\bf r}_j)$ is another optimizable degree of freedom which mostly affects the {\em nodal structure} of the wave-function. We make use of the short-ranged parametrization implemented in \cite{kim2018qmcpack}.

\begin{figure}[tb]
\includegraphics[scale=1]{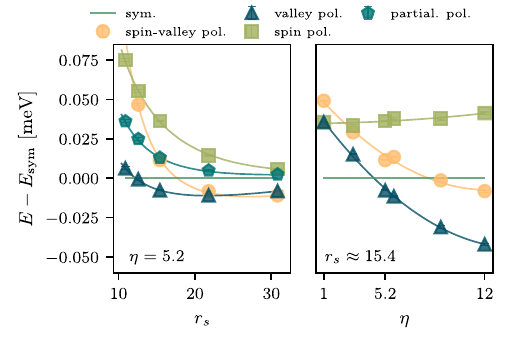}
\caption{The variational energies for states with different isospin polarization as a function of $r_s$ at selected anisotropy ($\eta=5.2$) (left) and as a function of $\eta$ at selected $r_s=15.4$ (right). 
}
\label{fig:fig2}
\end{figure}

Our VMC simulations include up to $N=176$ electrons, and are carried out using the package {\em qmcpack} \cite{kim2018qmcpack} with appropriate modifications. 
 There are two main sources of finite size errors: First, a discretization of the momenta ${\bf k}_j$ leads to discrete steps in the {\em kinetic energy} when the filling of the orbitals is determined. This can be alleviated via the introduction of twisted boundary conditions: We make use of the special twist method of Ref~\cite{dagrada2016exact}. Second, finite-size corrections to the potential pertain to how the {\em total energy} scales with the number of electrons. For the present case of screened interactions we resort to a phenomenological extrapolation to the thermodynamic limit, as detailed in \cite{SI}.

\parspace

{\it Phase diagram.} The VMC phase diagram of the metallic phases of Hamiltonian (\ref{eq:H}) as a function of $r_s$ and anisotropy $\eta$ is shown in Fig.~\ref{fig:fig1}. 
We simulate the regime $r_s \le 31$, but note that the AlAs experiment already observes a transition to an insulating phase at $r_s \approx 27$~\cite{hossain2022anisotropic}.
We find three stable phases of different isospin polarization for the considered range of densities and anisotropies. The unpolarized symmetric state is stable at large densities, whereas at very low denisities, the ground state is fully spin and valley polarized (SVP).  
The key feature, however, lies in the emergence of an intermediate, valley-polarized and spin-unpolarized  (VP) phase for $\eta  \gtrsim 3.5$. The density range where the VP phase is %stabilized 
stable further increases with the anisotropy of the system: For the largest simulated anisotropy ($\eta=12$), the VP state spans the whole range of simulated densities.

Simulations were performed for electron fluids with different isospin polarization on a discrete grid in the $r_s-\eta$ plane and the phase boundaries estimated by fitting the energies on horizontal and vertical lines \cite{SI}. The difference of the variational energies with respect to the symmetric state are displayed in Fig.~\ref{fig:fig2} for $\eta\approx 5.2$ and $r_{\ssm s}\approx 15.4$, respectively. We observe that of all the simulated polarization patterns, only the VP and the SVP states are ever the lowest energy states. Both a $3/4$-polarized state (where three of the four isospin-combinations are filled, with the same density for each of the isospin-combinations) and the spin-polarized state (SP) where both valleys are filled symmetrically are always higher in energy.  We have not calculated the energies of partially polarized phases, other than the $3/4$ state.
We shall comment further on the stability of the SVP phase below.

\begin{figure}[tb]
\includegraphics[scale=1]{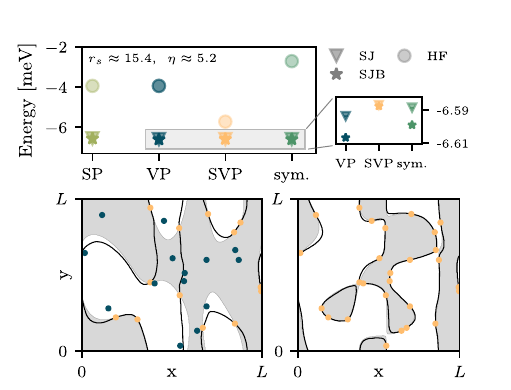}
\caption{Upper panel: The variational energies are plotted using Hartree-Fock (HF), a Slater-Jastrow (SJ) wave-function and Slater-Jastrow-Backflow (SJB) for selected density and anisotropy. The inset shows only VMC energies, in order to resolve the energy differences. We note that the inset is still too small to resolve the energy difference between a SJ and SJB ansatz for the SVP state. Lower panel: The nodal structures for a VP state (left) and SVP state (right) with 28 electrons. The nodes of the bare Slater determinant are depicted as boundaries between grey and white areas, whereas the black solid lines indicate the nodes of the Slater determinant with optimized Backflow transformation. The nodal structure is obtained by moving one electron through the $L\times L$ simulation cell while keeping the position of the other electrons fixed. The positions of the electrons with the same (opposite) isospin as the particle are shown as orange (blue) dots.}
\label{fig:fig3}
\end{figure}

Figure~\ref{fig:fig2} further demonstrates the growth of the energy difference between states with and without valley polarization as a function of increasing anisotropy.
Intuitively, this behaviour can  be understood as resulting from a correlation effect that penalizes states without full valley-polarization: Electrons from different valleys have different dispersions, which may hinder efficiently avoiding each other. The non-trivial nature of this effect becomes particularly apparent in comparison to mean-field theory. 

{\it Comparison to Hartree-Fock theory.} Within mean-field theory correlation effects cannot be captured, resulting in an accidental degeneracy of spin-and valley polarization (see Fig.~\ref{fig:fig3}) \cite{SI}.
Hartree-Fock theory faces a second shortcoming in the considered strongly interacting density regime. In particular, the phase diagram is determined by subtle correlation effects outside of the reach of mean-field theory. An exemplary case is illustrated in Fig.~\ref{fig:fig3} for $\eta=5.2$, $r_s=15.4$: While VMC results indicate the prevalence of a VP ground state, Hartree-Fock calculations \cite{SI} predict a SVP ground state, showing the well-known tendency of Hartree-Fock to over-favor symmetry-broken states. In addition, the energy differences between the states of different isospin polarization are two orders of magnitude smaller than within Hartree-Fock, demonstrating the relevance of subtle correlation effects.  %{\color{red} We might want to say that this shows the well known tendency of HF to over-favor broken symmetry states.}

Figure~\ref{fig:fig3} further shows that the energy gain resulting from introducing a Jastrow factor (correlation hole) is strongly isospin-polarization-dependent. In particular, the gain is smallest for the SVP state. The same observation can be made for the energy gain resulting from an optimizable nodal structure by introducing Backflow transformations; the gain is smallest for the SVP wave-function.
We can understand both observations by considering the nodal structure of the trial wave-functions, i.e. the positions in real space at which $\Psi_T({\bf r}_1, .. {\bf r}_N)=0$.
The nodal structure is solely determined by the determinantal part. Comparing the nodes of a bare Slater determinant (corresponding to a Hartree-Fock ansatz) and a Slater determinant with optimized Backflow transformations is a useful step to understand the validity of a mean-field approach.

The results are depicted in Fig.~\ref{fig:fig3} for the VP as well as the SVP state, showing the nodal structure of a two-dimensional ``slice'': It is obtained by fixing the position of all electrons but one and moving the latter through the simulation cell. As illustrated, Backflow transformations induce a much larger change in the nodal structure between different isospin flavours, as electrons with the same isospin are already kept apart by Pauli exclusion \cite{drummond2009phase, marchi2009correlation}. A similar reasoning might also be responsible for the polarization-dependent effect of the Jastrow factor.

These considerations are of particular relevance in connection to the stability of the SVP phase.
Concretely, one can e.g. obtain a more accurate results by including 3-body terms into the Jastrow factor \cite{kwon1993effects} or by performing diffusion Monte Carlo calculations \cite{foulkes2001quantum, varsano2001spin, drummond2009phase}. Following the above reasoning, when employing one of these options we expect a more enhanced effect on the energy of states with more than one isospin flavour. Consequently, the transition to the SVP state is likely to shift to larger $r_s$ upon using more accurate methods \cite{varsano2001spin}. The results on the two-dimensional {\it isotropic} electron gas \cite{rapisarda1996diffusion, drummond2009phase, marchi2009correlation} suggest that the transition may even be shifted to a density regime where Wigner crystallization becomes dominant. Whether a SVP phase exists for the considered model before entering the Wigner crystal is currently unclear.

\begin{figure}[tb]
\includegraphics[scale=1]{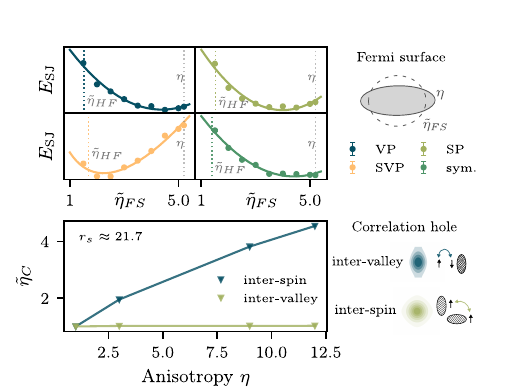}
\caption{ Upper panel: Slater-Jastrow variational energies at $r_s\approx 21.7$, $\eta=5.2$ as a function of effective Fermi surface anisotropy $\tilde{\eta}_{\ssm FS}$. Backflow transformations are applied in the next step of the simulations only for the wave-functions with $\tilde{\eta}_{\ssm FS}$ in the vicinity of the minimum. The effective Hartree-Fock anisotropy $\tilde{\eta}_{\ssm HF}$  and the bare anisotropy $\eta$ are indicated via grey dotted lines. 
Lower panel: the anisotropy of the inter-isospin correlation hole for the VP state (blue) and the SP state (green) as a function of bare anisotropy.  On the right hand side of the plot, the Jastrow factor is plotted as an exemplary illustration for $\eta=9$ and $114$ electrons.  %{\color{red} [I don't know what exemplary plotted means]}
}
\label{fig:fig4}
\end{figure}

{\it Effective anisotropy}. 
The effective Fermi surface anisotropy $\tilde{\eta}_{\ssm FS}$ of the ground state has physical implications, e.g. on transport properties. We determine the effective anisotropy by scanning through a range $1\leq \tilde{\eta}_{\ssm FS} \leq \eta$ and picking the solution with lowest energy. Here, $\eta$ is the bare anisotropy of the system. Figure~\ref{fig:fig4} depicts the so-obtained variational energies as a function of $\tilde{\eta}_{\ssm FS}$ for different isospin polarizations. The position of the minimum can be understood as the effective anisotropy of the Fermi surface within VMC. We can compare this result with the renormalized anisotropy $\tilde{\eta}_{\ssm HF}$ obtained in Hartree-Fock calculations, also shown in Fig.~\ref{fig:fig4}. We observe that Hartree-Fock strongly underestimates the effective anisotropy for all polarizations except the SVP state: As detailed above, correlation effects beyond Hartree-Fock are expected to play a more dominant role for states with different isospin flavours, which explains the polarization-dependent discrepancy on the effective anisotropy obtained via Hartree-Fock and VMC.

In order to explore the interplay between anisotropy and correlation effects, we turn to a more concrete study of the correlation hole and the effective anisotropy $\tilde{\eta}_{\ssm C}$ thereof. 
Our objective here is to compare the correlation hole between two electrons of different (iso)spin flavour. The underlying motivation lies in acquiring intuition about the emergence of the energy gap between valley-polarized and valley-unpolarized states with increasing ansiotropy.
Direct access to the correlation hole is given in form of the two-body Jastrow factor. We obtain the effective ansiotropy $\tilde{\eta}_{\ssm C}$ by fitting an ellipse to the equipotential lines of the Jastrow factor and taking the mean over the three largest system sizes. 
The results are depicted in Fig.~\ref{fig:fig4} and illuminate the comparative energy gain that valley-polarization induces in the presence of anisotropy. While two electrons within the same valley are able to optimally avoid each other by forming an anisotropic correlation hole, the correlation hole in between valleys is isotropic: Intuitively one can deduce, that not being able to capitalize on the anisotropy of the system in the formation of a correlation hole will introduce an energy offset for valley-unpolarized states.

{\em Comparison to experiment.} At the anisotropy relevant for AlAs ($\eta\approx 5.2$) \cite{hossain2021spontaneous}, our VMC calculations predict a transition from a symmetric state to a stable valley-polarized phase in the relevant density regime, which is in agreement with the experimental observations. 
There is, however, a quantitative mismatch between the critical $r_s$ obtained numerically ($r_s \approx 12$) and in the experiment ($r_s \approx 20$). In order to understand the root of this mismatch, we turn to a more detailed discussion of the Hamiltonian~(\ref{eq:H}) and neglected terms therein. 

Concretely, we used symmetric gates at a distance of $d=100$ nm, whereas in the experiment there is only a single gate at a much larger distance. However, simulations at selected values of $d$ in the range $70$ nm $\leq d \leq 300$ nm did not indicate a significant shift of the phase boundaries. 

In addition, we have neglected several effects, including electron-phonon coupling, inter-valley scattering terms, and the finite thickness of the quantum well. In \cite{SI}, we estimate the typical electron-phonon coupling and the inter-valley scattering terms, and find that they are both small compared to the energy differences between states of different spin and valley polarization (Fig.~\ref{fig:fig2}). 
Potentially more significant is the thickness of the quantum well, which is $w\approx 20\,\rm{nm}$ in Ref.~\cite{hossain2021spontaneous}; this modifies the Coulomb interaction from $v(|\mathbf{r}|)\sim 1/r$ at $r\gg w$ to $\ln(w/r)$ at $r\ll w$. Taking this effect into account is expected to shift the transition between the symmetric and the valley-polarized states to larger $r_s$, closer to the experimental value (since increasing $w$ weakens the Coulomb interaction at short distances).

Another possible explanation of the apparent quantitative mismatch in the critical $r_s$ is an uncertainty in the anisotropy of AlAs: In particular, a smaller anisotropy than $\eta\approx 5.2$ results in a larger critical $r_s$.  

At even lower densities ($r_s \approx 27 $), we find a transition to a SVP metallic state. The existence and stability of this phase is, however, as detailed above, unclear.
Experimentally a preference for spin-valley polarization has been found at $r_s \approx 35 $, but the observed phase is insulating rather than metallic. For more conclusive statements about this phase, numerical consideration of insulating states would be required.

In addition, we note that according to Ref.~\cite{shchepetilnikov2020spin} the influence of spin-orbit coupling might be non-neglibible in AlAs. We leave its consideration to future studies.

{\it Perspectives} 
We explored the low-density phase diagram of an anisotropic multi-valley electron gas with VMC using a Slater-Jastrow-Backflow wave-function.
In contrast to the isotropic case, we find that there is a strong evidence for the existence of a symmetry-broken {\em metallic} phase. In particular, as the anisotropy increases, a stable valley-polarized, spin-unpolarized phase emerges. This result is in qualitative agreement with recent experiments on AlAs, where a spontaneous transition to a nematic valley-polarized state is observed at $r_s\gtrsim 20$ \cite{hossain2021spontaneous}.
While we find this transition at higher densities ($r_s\gtrsim 12$), we identify the finite thickness of the well (neglected in our calculation) as the most probable source of this difference. 

Here, we focused on spin or valley polarized, metallic phases: Study of other phases such as inter-valley coherent order \cite{chatterjee2022inter} as well as crystalline phases \cite{wigner1934interaction, hossain2022anisotropic, calvera2022pseudo} pose an interesting avenue for future research.
In addition, more accurate methods such as diffusion Monte Carlo \cite{foulkes2001quantum}, auxiliary-field quantum Monte Carlo \cite{zhang2003quantum, zhang201315} or other classes of wave-functions \cite{pescia2023message, cassella2023discovering, wilson2022wave, li2022ab} may reveal a quantitative change in behaviour and yield further physical insight.
Comparison to mean-field calculations reveal the relevance of correlation effect and call of strongly correlated methods as the one presented here.

\vspace{0.5cm}

\begin{acknowledgments}
We thank Sankar Das Sarma and Mansour Shayegan for their comments on this manuscript. 
This work has received funding from the
European Research Council under grant agreement no. 771503. S.D.H. acknowledges support by the Benoziyo Endowment Fund for the Advancement of Science. S.A.K and E.B. were supported by NSF-BSF award DMR-2000987. E.B. acknowledges support from the European Research Council (ERC) under grant HQMAT (grant agreement No. 817799). E.B. thanks the hospitality of the Aspen Center for Physics, supported by National Science Foundation grant PHY-2210452, where part of this work was done.
\end{acknowledgments}

\bibliographystyle{unsrt}
\bibliography{papers.bib}

\begin{thebibliography}{10}

\bibitem{ando1982electronic}
Tsuneya Ando, Alan~B Fowler, and Frank Stern.
\newblock Electronic properties of two-dimensional systems.
\newblock {\em Reviews of Modern Physics}, 54(2):437, 1982.

\bibitem{tanatar1989ground}
B~Tanatar and David~M Ceperley.
\newblock Ground state of the two-dimensional electron gas.
\newblock {\em Physical Review B}, 39(8):5005, 1989.

\bibitem{kwon1993effects}
Yongkyung Kwon, DM~Ceperley, and Richard~M Martin.
\newblock Effects of three-body and backflow correlations in the
  two-dimensional electron gas.
\newblock {\em Physical Review B}, 48(16):12037, 1993.

\bibitem{rapisarda1996diffusion}
Francesco Rapisarda and Gaetano Senatore.
\newblock Diffusion monte carlo study of electrons in two-dimensional layers.
\newblock {\em Australian journal of physics}, 49(1):161--182, 1996.

\bibitem{attaccalite2002correlation}
Claudio Attaccalite, Saverio Moroni, Paola Gori-Giorgi, and Giovanni~B
  Bachelet.
\newblock Correlation energy and spin polarization in the 2d electron gas.
\newblock {\em Physical review letters}, 88(25):256601, 2002.

\bibitem{gori2003two}
Paola Gori-Giorgi, Claudio Attaccalite, Saverio Moroni, and Giovanni~B
  Bachelet.
\newblock Two-dimensional electron gas: Correlation energy versus density and
  spin polarization.
\newblock {\em International journal of quantum chemistry}, 91(2):126--130,
  2003.

\bibitem{drummond2009phase}
ND~Drummond and RJ~Needs.
\newblock Phase diagram of the low-density two-dimensional homogeneous electron
  gas.
\newblock {\em Physical review letters}, 102(12):126402, 2009.

\bibitem{marchi2009correlation}
M~Marchi, S~De~Palo, S~Moroni, and Gaetano Senatore.
\newblock Correlation energy and spin susceptibility of a two-valley
  two-dimensional electron gas.
\newblock {\em Physical Review B}, 80(3):035103, 2009.

\bibitem{spivak2004phases}
Boris Spivak and Steven~A Kivelson.
\newblock Phases intermediate between a two-dimensional electron liquid and
  wigner crystal.
\newblock {\em Physical Review B}, 70(15):155114, 2004.

\bibitem{jamei2005universal}
Reza Jamei, Steven Kivelson, and Boris Spivak.
\newblock Universal aspects of coulomb-frustrated phase separation.
\newblock {\em Physical review letters}, 94(5):056805, 2005.

\bibitem{raghu2011superconductivity}
S~Raghu and SA~Kivelson.
\newblock Superconductivity from repulsive interactions in the two-dimensional
  electron gas.
\newblock {\em Physical Review B}, 83(9):094518, 2011.

\bibitem{kim2022interstitial}
Kyung-Su Kim, Chaitanya Murthy, Akshat Pandey, Steven~A Kivelson, et~al.
\newblock Interstitial-induced ferromagnetism in a two-dimensional wigner
  crystal.
\newblock {\em Physical Review Letters}, 129(22):227202, 2022.

\bibitem{loos2016uniform}
Pierre-Fran{\c{c}}ois Loos and Peter~MW Gill.
\newblock The uniform electron gas.
\newblock {\em Wiley Interdisciplinary Reviews: Computational Molecular
  Science}, 6(4):410--429, 2016.

\bibitem{zhou2021superconductivity}
Haoxin Zhou, Tian Xie, Takashi Taniguchi, Kenji Watanabe, and Andrea~F Young.
\newblock Superconductivity in rhombohedral trilayer graphene.
\newblock {\em Nature}, 598(7881):434--438, 2021.

\bibitem{seiler2022quantum}
Anna~M Seiler, Fabian~R Geisenhof, Felix Winterer, Kenji Watanabe, Takashi
  Taniguchi, Tianyi Xu, Fan Zhang, and R~Thomas Weitz.
\newblock Quantum cascade of correlated phases in trigonally warped bilayer
  graphene.
\newblock {\em Nature}, 608(7922):298--302, 2022.

\bibitem{cao2018unconventional}
Yuan Cao, Valla Fatemi, Shiang Fang, Kenji Watanabe, Takashi Taniguchi,
  Efthimios Kaxiras, and Pablo Jarillo-Herrero.
\newblock Unconventional superconductivity in magic-angle graphene
  superlattices.
\newblock {\em Nature}, 556(7699):43--50, 2018.

\bibitem{chen2019signatures}
Guorui Chen, Aaron~L Sharpe, Patrick Gallagher, Ilan~T Rosen, Eli~J Fox, Lili
  Jiang, Bosai Lyu, Hongyuan Li, Kenji Watanabe, Takashi Taniguchi, et~al.
\newblock Signatures of tunable superconductivity in a trilayer graphene
  moir{\'e} superlattice.
\newblock {\em Nature}, 572(7768):215--219, 2019.

\bibitem{park2021tunable}
Jeong~Min Park, Yuan Cao, Kenji Watanabe, Takashi Taniguchi, and Pablo
  Jarillo-Herrero.
\newblock Tunable strongly coupled superconductivity in magic-angle twisted
  trilayer graphene.
\newblock {\em Nature}, 590(7845):249--255, 2021.

\bibitem{bistritzer2011moire}
Rafi Bistritzer and Allan~H MacDonald.
\newblock Moir{\'e} bands in twisted double-layer graphene.
\newblock {\em Proceedings of the National Academy of Sciences},
  108(30):12233--12237, 2011.

\bibitem{cao2018correlated}
Yuan Cao, Valla Fatemi, Ahmet Demir, Shiang Fang, Spencer~L Tomarken, Jason~Y
  Luo, Javier~D Sanchez-Yamagishi, Kenji Watanabe, Takashi Taniguchi, Efthimios
  Kaxiras, et~al.
\newblock Correlated insulator behaviour at half-filling in magic-angle
  graphene superlattices.
\newblock {\em Nature}, 556(7699):80--84, 2018.

\bibitem{shen2020correlated}
Cheng Shen, Yanbang Chu, QuanSheng Wu, Na~Li, Shuopei Wang, Yanchong Zhao, Jian
  Tang, Jieying Liu, Jinpeng Tian, Kenji Watanabe, et~al.
\newblock Correlated states in twisted double bilayer graphene.
\newblock {\em Nature Physics}, 16(5):520--525, 2020.

\bibitem{zhao2023gate}
Wenjin Zhao, Bowen Shen, Zui Tao, Zhongdong Han, Kaifei Kang, Kenji Watanabe,
  Takashi Taniguchi, Kin~Fai Mak, and Jie Shan.
\newblock Gate-tunable heavy fermions in a moir{\'e} kondo lattice.
\newblock {\em Nature}, 616(7955):61--65, 2023.

\bibitem{mak2022semiconductor}
Kin~Fai Mak and Jie Shan.
\newblock Semiconductor moir{\'e} materials.
\newblock {\em Nature Nanotechnology}, 17(7):686--695, 2022.

\bibitem{hofmann2022fermionic}
Johannes~S Hofmann, Eslam Khalaf, Ashvin Vishwanath, Erez Berg, and Jong~Yeon
  Lee.
\newblock Fermionic monte carlo study of a realistic model of twisted bilayer
  graphene.
\newblock {\em Physical Review X}, 12(1):011061, 2022.

\bibitem{Yang2023Metal}
Yubo Yang, Miguel Morales, and Shiwei Zhang.
\newblock Metal-insulator transition in transition metal dichalcogenide
  heterobilayer: accurate treatment of interaction.
\newblock {\em arXiv preprint arXiv:2306.14954}, 2023.

\bibitem{shayegan2006two}
M~Shayegan, EP~De~Poortere, O~Gunawan, YP~Shkolnikov, E~Tutuc, and K~Vakili.
\newblock Two-dimensional electrons occupying multiple valleys in alas.
\newblock {\em physica status solidi (b)}, 243(14):3629--3642, 2006.

\bibitem{hossain2020observation}
Md~S Hossain, MK~Ma, KA~Rosales, YJ~Chung, LN~Pfeiffer, KW~West, KW~Baldwin,
  and Mansour Shayegan.
\newblock Observation of spontaneous ferromagnetism in a two-dimensional
  electron system.
\newblock {\em Proceedings of the National Academy of Sciences},
  117(51):32244--32250, 2020.

\bibitem{hossain2021spontaneous}
Md~S Hossain, MK~Ma, KA~Villegas-Rosales, YJ~Chung, LN~Pfeiffer, KW~West,
  KW~Baldwin, and M~Shayegan.
\newblock Spontaneous valley polarization of itinerant electrons.
\newblock {\em Physical review letters}, 127(11):116601, 2021.

\bibitem{hossain2022anisotropic}
Md~S Hossain, MK~Ma, KA~Villegas-Rosales, YJ~Chung, LN~Pfeiffer, KW~West,
  KW~Baldwin, and M~Shayegan.
\newblock Anisotropic two-dimensional disordered wigner solid.
\newblock {\em Physical review letters}, 129(3):036601, 2022.

\bibitem{footnote1}
{$r_s=m^*e^2/\epsilon\hbar^2\sqrt{\pi n}$}, where $m^*$ is the effective band
  mass, $\epsilon$ the dielectric constant, and $n$ and $e$ the electron
  density and charge, respectively.

\bibitem{Ahn2022}
Seongjin Ahn and Sankar Das~Sarma.
\newblock Valley polarization transition in a two-dimensional electron gas.
\newblock {\em Phys. Rev. B}, 105:L241411, Jun 2022.

\bibitem{SI}
Supplemental material.

\bibitem{throckmorton2012fermions}
Robert~E Throckmorton and Oskar Vafek.
\newblock Fermions on bilayer graphene: Symmetry breaking for b= 0 and $\nu$=
  0.
\newblock {\em Physical Review B}, 86(11):115447, 2012.

\bibitem{jastrow1955many}
Robert Jastrow.
\newblock Many-body problem with strong forces.
\newblock {\em Physical Review}, 98(5):1479, 1955.

\bibitem{lee1981green}
Michael~A Lee, KE~Schmidt, MH~Kalos, and GV~Chester.
\newblock Green's function monte carlo method for liquid he 3.
\newblock {\em Physical Review Letters}, 46(11):728, 1981.

\bibitem{kwon1998effects}
Yongkyung Kwon, DM~Ceperley, and Richard~M Martin.
\newblock Effects of backflow correlation in the three-dimensional electron
  gas: Quantum monte carlo study.
\newblock {\em Physical Review B}, 58(11):6800, 1998.

\bibitem{sorella1998green}
Sandro Sorella.
\newblock Green function monte carlo with stochastic reconfiguration.
\newblock {\em Physical review letters}, 80(20):4558, 1998.

\bibitem{sorella2007weak}
Sandro Sorella, Michele Casula, and Dario Rocca.
\newblock Weak binding between two aromatic rings: Feeling the van der waals
  attraction by quantum monte carlo methods.
\newblock {\em The Journal of chemical physics}, 127(1):014105, 2007.

\bibitem{ahn2021anisotropic}
Seongjin Ahn and S~Das Sarma.
\newblock Anisotropic fermionic quasiparticles.
\newblock {\em Physical Review B}, 103(4):045303, 2021.

\bibitem{whitehead2016jastrow}
TM~Whitehead, MH~Michael, and GJ~Conduit.
\newblock Jastrow correlation factor for periodic systems.
\newblock {\em Physical Review B}, 94(3):035157, 2016.

\bibitem{kim2018qmcpack}
Jeongnim Kim, Andrew~D Baczewski, Todd~D Beaudet, Anouar Benali, M~Chandler
  Bennett, Mark~A Berrill, Nick~S Blunt, Edgar Josu{\'e}~Landinez Borda,
  Michele Casula, David~M Ceperley, et~al.
\newblock Qmcpack: an open source ab initio quantum monte carlo package for the
  electronic structure of atoms, molecules and solids.
\newblock {\em Journal of Physics: Condensed Matter}, 30(19):195901, 2018.

\bibitem{dagrada2016exact}
Mario Dagrada, Seher Karakuzu, Ver{\'o}nica~Laura Vildosola, Michele Casula,
  and Sandro Sorella.
\newblock Exact special twist method for quantum monte carlo simulations.
\newblock {\em Physical Review B}, 94(24):245108, 2016.

\bibitem{foulkes2001quantum}
WMC Foulkes, Lubos Mitas, RJ~Needs, and Guna Rajagopal.
\newblock Quantum monte carlo simulations of solids.
\newblock {\em Reviews of Modern Physics}, 73(1):33, 2001.

\bibitem{varsano2001spin}
D~Varsano, S~Moroni, and Gaetano Senatore.
\newblock Spin-polarization transition in the two-dimensional electron gas.
\newblock {\em Europhysics Letters}, 53(3):348, 2001.

\bibitem{shchepetilnikov2020spin}
AV~Shchepetilnikov, Alina~R Khisameeva, Yu~A Nefyodov, Igor~V Kukushkin, Lars
  Tiemann, Christian Reichl, Werner Dietsche, and Werner Wegscheider.
\newblock Spin-orbit interaction in alas quantum wells.
\newblock {\em Physica E: Low-dimensional Systems and Nanostructures},
  124:114278, 2020.

\bibitem{chatterjee2022inter}
Shubhayu Chatterjee, Taige Wang, Erez Berg, and Michael~P Zaletel.
\newblock Inter-valley coherent order and isospin fluctuation mediated
  superconductivity in rhombohedral trilayer graphene.
\newblock {\em Nature Communications}, 13(1):6013, 2022.

\bibitem{wigner1934interaction}
Eugene Wigner.
\newblock On the interaction of electrons in metals.
\newblock {\em Physical Review}, 46(11):1002, 1934.

\bibitem{calvera2022pseudo}
Vladimir Calvera, Steven~A Kivelson, and Erez Berg.
\newblock Pseudo-spin order of wigner crystals in multi-valley electron gases.
\newblock {\em arXiv preprint arXiv:2210.09326}, 2022.

\bibitem{zhang2003quantum}
Shiwei Zhang and Henry Krakauer.
\newblock Quantum monte carlo method using phase-free random walks with slater
  determinants.
\newblock {\em Physical review letters}, 90(13):136401, 2003.

\bibitem{zhang201315}
Shiwei Zhang.
\newblock 15 auxiliary-field quantum monte carlo for correlated electron
  systems.
\newblock {\em Emergent Phenomena in Correlated Matter}, 2013.

\bibitem{pescia2023message}
Gabriel Pescia, Jannes Nys, Jane Kim, Alessandro Lovato, and Giuseppe Carleo.
\newblock Message-passing neural quantum states for the homogeneous electron
  gas.
\newblock {\em arXiv preprint arXiv:2305.07240}, 2023.

\bibitem{cassella2023discovering}
Gino Cassella, Halvard Sutterud, Sam Azadi, ND~Drummond, David Pfau, James~S
  Spencer, and W~Matthew~C Foulkes.
\newblock Discovering quantum phase transitions with fermionic neural networks.
\newblock {\em Physical Review Letters}, 130(3):036401, 2023.

\bibitem{wilson2022wave}
Max Wilson, Saverio Moroni, Markus Holzmann, Nicholas Gao, Filip Wudarski, Tejs
  Vegge, and Arghya Bhowmik.
\newblock Wave function ansatz (but periodic) networks and the homogeneous
  electron gas.
\newblock {\em arXiv preprint arXiv:2202.04622}, 2022.

\bibitem{li2022ab}
Xiang Li, Zhe Li, and Ji~Chen.
\newblock Ab initio calculation of real solids via neural network ansatz.
\newblock {\em Nature Communications}, 13(1):7895, 2022.

\end{thebibliography}


\begin{thebibliography}{10}

\bibitem{hossain2021spontaneous}
Md~S Hossain, MK~Ma, KA~Villegas-Rosales, YJ~Chung, LN~Pfeiffer, KW~West,
  KW~Baldwin, and M~Shayegan.
\newblock Spontaneous valley polarization of itinerant electrons.
\newblock {\em Physical review letters}, 127(11):116601, 2021.

\bibitem{de2005effects}
S~De~Palo, M~Botti, S~Moroni, and Gaetano Senatore.
\newblock Effects of thickness on the spin susceptibility of the two
  dimensional electron gas.
\newblock {\em Physical review letters}, 94(22):226405, 2005.

\bibitem{gold1987electronic}
A~Gold.
\newblock Electronic transport properties of a two-dimensional electron gas in
  a silicon quantum-well structure at low temperature.
\newblock {\em Physical Review B}, 35(2):723, 1987.

\bibitem{adachi1985gaas}
Sadao Adachi.
\newblock {GaAs, AlAs, and Al$_x$ Ga$_{1-x}$ As: Material parameters for use in
  research and device applications}.
\newblock {\em Journal of applied physics}, 58(3):R1--R29, 1985.

\bibitem{Charbonneau1991MeasurementElasticAlAs}
S.~Charbonneau, Jeff~F. Young, P.~T. Coleridge, and B.~Kettles.
\newblock Experimental determination of the ${\mathit{x}}_{6}$ shear tetragonal
  deformation potential of alas.
\newblock {\em Phys. Rev. B}, 44:8312--8314, Oct 1991.

\bibitem{Gunawan2006ValleySusceptibility}
O.~Gunawan, Y.~P. Shkolnikov, K.~Vakili, T.~Gokmen, E.~P. De~Poortere, and
  M.~Shayegan.
\newblock Valley susceptibility of an interacting two-dimensional electron
  system.
\newblock {\em Phys. Rev. Lett.}, 97:186404, Nov 2006.

\bibitem{casula2005new}
Michele Casula et~al.
\newblock New qmc approaches for the simulation of electronic systems: a first
  application to aromatic molecules and transition metal compounds.
\newblock 2005.

\bibitem{sorella1998green}
Sandro Sorella.
\newblock Green function monte carlo with stochastic reconfiguration.
\newblock {\em Physical review letters}, 80(20):4558, 1998.

\bibitem{park2020geometry}
Chae-Yeun Park and Michael~J Kastoryano.
\newblock Geometry of learning neural quantum states.
\newblock {\em Physical Review Research}, 2(2):023232, 2020.

\bibitem{pfau2020ab}
David Pfau, James~S Spencer, Alexander~GDG Matthews, and W~Matthew~C Foulkes.
\newblock Ab initio solution of the many-electron schr{\"o}dinger equation with
  deep neural networks.
\newblock {\em Physical Review Research}, 2(3):033429, 2020.

\bibitem{foulkesvariational}
WMC Foulkes.
\newblock Variational wave functions for molecules and solids.

\bibitem{bengio2017deep}
Yoshua Bengio, Ian Goodfellow, and Aaron Courville.
\newblock {\em Deep learning}, volume~1.
\newblock MIT press Cambridge, MA, USA, 2017.

\bibitem{cassella2023discovering}
Gino Cassella, Halvard Sutterud, Sam Azadi, ND~Drummond, David Pfau, James~S
  Spencer, and W~Matthew~C Foulkes.
\newblock Discovering quantum phase transitions with fermionic neural networks.
\newblock {\em Physical Review Letters}, 130(3):036401, 2023.

\bibitem{wilson2022wave}
Max Wilson, Saverio Moroni, Markus Holzmann, Nicholas Gao, Filip Wudarski, Tejs
  Vegge, and Arghya Bhowmik.
\newblock Wave function ansatz (but periodic) networks and the homogeneous
  electron gas.
\newblock {\em arXiv preprint arXiv:2202.04622}, 2022.

\bibitem{li2022ab}
Xiang Li, Zhe Li, and Ji~Chen.
\newblock Ab initio calculation of real solids via neural network ansatz.
\newblock {\em Nature Communications}, 13(1):7895, 2022.

\bibitem{pescia2023message}
Gabriel Pescia, Jannes Nys, Jane Kim, Alessandro Lovato, and Giuseppe Carleo.
\newblock Message-passing neural quantum states for the homogeneous electron
  gas.
\newblock {\em arXiv preprint arXiv:2305.07240}, 2023.

\bibitem{jastrow1955many}
Robert Jastrow.
\newblock Many-body problem with strong forces.
\newblock {\em Physical Review}, 98(5):1479, 1955.

\bibitem{holzmann2016theory}
Markus Holzmann, Raymond~C Clay~III, Miguel~A Morales, Norm~M Tubman, David~M
  Ceperley, and Carlo Pierleoni.
\newblock Theory of finite size effects for electronic quantum monte carlo
  calculations of liquids and solids.
\newblock {\em Physical Review B}, 94(3):035126, 2016.

\bibitem{kim2018qmcpack}
Jeongnim Kim, Andrew~D Baczewski, Todd~D Beaudet, Anouar Benali, M~Chandler
  Bennett, Mark~A Berrill, Nick~S Blunt, Edgar Josu{\'e}~Landinez Borda,
  Michele Casula, David~M Ceperley, et~al.
\newblock Qmcpack: an open source ab initio quantum monte carlo package for the
  electronic structure of atoms, molecules and solids.
\newblock {\em Journal of Physics: Condensed Matter}, 30(19):195901, 2018.

\bibitem{ceperley1978ground}
D~Ceperley.
\newblock Ground state of the fermion one-component plasma: A monte carlo study
  in two and three dimensions.
\newblock {\em Physical Review B}, 18(7):3126, 1978.

\bibitem{kwon1993effects}
Yongkyung Kwon, DM~Ceperley, and Richard~M Martin.
\newblock Effects of three-body and backflow correlations in the
  two-dimensional electron gas.
\newblock {\em Physical Review B}, 48(16):12037, 1993.

\bibitem{drummond2004jastrow}
Neil~D Drummond, Michael~D Towler, and RJ~Needs.
\newblock Jastrow correlation factor for atoms, molecules, and solids.
\newblock {\em Physical Review B}, 70(23):235119, 2004.

\bibitem{drummond2009phase}
ND~Drummond and RJ~Needs.
\newblock Phase diagram of the low-density two-dimensional homogeneous electron
  gas.
\newblock {\em Physical review letters}, 102(12):126402, 2009.

\bibitem{whitehead2016jastrow}
TM~Whitehead, MH~Michael, and GJ~Conduit.
\newblock Jastrow correlation factor for periodic systems.
\newblock {\em Physical Review B}, 94(3):035157, 2016.

\bibitem{rios2006inhomogeneous}
P~L{\'o}pez R{\'\i}os, Ao~Ma, Neil~D Drummond, Michael~D Towler, and Richard~J
  Needs.
\newblock Inhomogeneous backflow transformations in quantum monte carlo
  calculations.
\newblock {\em Physical Review E}, 74(6):066701, 2006.

\bibitem{lee1981green}
Michael~A Lee, KE~Schmidt, MH~Kalos, and GV~Chester.
\newblock Green's function monte carlo method for liquid he 3.
\newblock {\em Physical Review Letters}, 46(11):728, 1981.

\bibitem{kwon1998effects}
Yongkyung Kwon, DM~Ceperley, and Richard~M Martin.
\newblock Effects of backflow correlation in the three-dimensional electron
  gas: Quantum monte carlo study.
\newblock {\em Physical Review B}, 58(11):6800, 1998.

\bibitem{schmidt1981structure}
KE~Schmidt, Michael~A Lee, MH~Kalos, and GV~Chester.
\newblock Structure of the ground state of a fermion fluid.
\newblock {\em Physical Review Letters}, 47(11):807, 1981.

\bibitem{sangster1976interionic}
MJL Sangster and M~Dixon.
\newblock Interionic potentials in alkali halides and their use in simulations
  of the molten salts.
\newblock {\em Advances in Physics}, 25(3):247--342, 1976.

\bibitem{toukmaji1996ewald}
Abdulnour~Y Toukmaji and John~A Board~Jr.
\newblock Ewald summation techniques in perspective: a survey.
\newblock {\em Computer physics communications}, 95(2-3):73--92, 1996.

\bibitem{throckmorton2012fermions}
Robert~E Throckmorton and Oskar Vafek.
\newblock Fermions on bilayer graphene: Symmetry breaking for b= 0 and $\nu$=
  0.
\newblock {\em Physical Review B}, 86(11):115447, 2012.

\bibitem{goodwin2020critical}
Zachary~AH Goodwin, Valerio Vitale, Fabiano Corsetti, Dmitri~K Efetov, Arash~A
  Mostofi, and Johannes Lischner.
\newblock Critical role of device geometry for the phase diagram of twisted
  bilayer graphene.
\newblock {\em Physical Review B}, 101(16):165110, 2020.

\bibitem{wolframurl}
{Wolfram Functions}.
\newblock \url{http://functions.wolfram.com/09.04.06.0027.01}.

\bibitem{drummond2008finite}
ND~Drummond, RJ~Needs, A~Sorouri, and WMC Foulkes.
\newblock Finite-size errors in continuum quantum monte carlo calculations.
\newblock {\em Physical Review B}, 78(12):125106, 2008.

\bibitem{kwee2008finite}
Hendra Kwee, Shiwei Zhang, and Henry Krakauer.
\newblock Finite-size correction in many-body electronic structure
  calculations.
\newblock {\em Physical review letters}, 100(12):126404, 2008.

\bibitem{ceperley1980ground}
David~M Ceperley and Berni~J Alder.
\newblock Ground state of the electron gas by a stochastic method.
\newblock {\em Physical review letters}, 45(7):566, 1980.

\bibitem{lin2001twist}
C~Lin, FH~Zong, and David~M Ceperley.
\newblock Twist-averaged boundary conditions in continuum quantum monte carlo
  algorithms.
\newblock {\em Physical Review E}, 64(1):016702, 2001.

\bibitem{dagrada2016exact}
Mario Dagrada, Seher Karakuzu, Ver{\'o}nica~Laura Vildosola, Michele Casula,
  and Sandro Sorella.
\newblock Exact special twist method for quantum monte carlo simulations.
\newblock {\em Physical Review B}, 94(24):245108, 2016.

\bibitem{chiesa2006finite}
Simone Chiesa, David~M Ceperley, Richard~M Martin, and Markus Holzmann.
\newblock Finite-size error in many-body simulations with long-range
  interactions.
\newblock {\em Physical review letters}, 97(7):076404, 2006.

\bibitem{giuliani2005quantum}
Gabriele Giuliani and Giovanni Vignale.
\newblock {\em Quantum theory of the electron liquid}.
\newblock Cambridge university press, 2005.

\bibitem{rapisarda1996diffusion}
Francesco Rapisarda and Gaetano Senatore.
\newblock Diffusion monte carlo study of electrons in two-dimensional layers.
\newblock {\em Australian journal of physics}, 49(1):161--182, 1996.

\end{thebibliography}

\end{document}

%TC:ignore
\widetext

\date{\today}
\author{Agnes Valenti}
\affiliation{Institute for Theoretical Physics, ETH Zurich, CH-8093, Switzerland}
\author{Vladimir Calvera}
\affiliation{Department of Physics, Stanford University, Stanford, CA 94305, USA}
\author{Steven A. Kivelson}
\affiliation{Department of Physics, Stanford University, Stanford, CA 94305, USA}
\author{Erez Berg}
\affiliation{Department of Condensed Matter Physics, Weizmann Institute of Science, Rehovot 76001, Israel}
\author{Sebastian D. Huber}
\affiliation{Institute for Theoretical Physics, ETH Zurich, CH-8093, Switzerland}

\title{Supplemental Material for ``Nematic metal in a multi-valley electron gas: \\ Variational Monte Carlo analysis and application to AlAs''}

\maketitle

\tableofcontents

\setcounter{equation}{0}
\setcounter{figure}{0}
\setcounter{page}{1}
\makeatletter
\renewcommand{\theequation}{S\arabic{equation}}
\renewcommand{\thefigure}{S\arabic{figure}}
\renewcommand{\bibnumfmt}[1]{[S#1]}
\renewcommand{\citenumfont}[1]{S#1}

\section{Approximations in the model Hamiltonian}
In our VMC simulations, we consider the Hamiltonian
\begin{align}
\label{eq:H}
H&=-\sum \limits_{i} \frac{1}{2m^{*}}\big(\eta^{\tau_i /2} \partial_{i,x}^2+ \eta^{-\tau_i /2} \partial_{i,y}^2 \big) + \sum \limits_{i<j} V(|{\bf r}_i- {\bf r}_j|), \\
V(|{\bf r}_i-{\bf r}_j|)&=\frac{1}{(2\pi)^2}\int {\rm d} {\bf q}\, {\rm e}^{i{\bf q} |{\bf r}_i-{\bf r}_j|}v({\bf q}), \\
v({\bf q})&= \frac{e^2}{2 \epsilon_0 \epsilon} \frac{\tanh (d |{\bf q}|)}{|{\bf q}|}.
\end{align}
where $m^{*}$ is the effective mass, $\eta$ denotes the anisotropy, the sum runs over all particles $i$ and $d $ corresponds to the gate-distance. 
In this section, we discuss neglected terms and made approximations in comparison to experiments on an AlAs quantum well \cite{hossain2021spontaneous}.

\subsection{Finite thickness}
A source that may affect the regions of stability in the phase diagram is the finite thickness of the AlAs quantum well. In~\cite{de2005effects} it has been shown, that while the finite thickness significantly alters the spin susceptibility, it only results in a slight shift of the phase boundaries.   

We neglect its effect here, but note that it still may be non-negligible as the results in~\cite{de2005effects} are obtained via perturbation theory and the estimation of phase boundaries is sensitive to small energy differences. In principle, finite-size thickness could be directly taken into account by performing simulations in a slab of finite thickness. This would, however, require increased computational effort. An estimation of the effects while keeping the simulations strictly two-dimensional can also be obtained via a simple approximation, using a device-specific form factor $F(q)$ that modifies the interaction $v(q)$ in Fourier space~\cite{de2005effects} to $\tilde{v}(q)=v(q)F(q)$, where $F(q=0)=1$ and $F(q\gg 1/w)\sim C/(wq)$ (here, $w$ is the width of the quantum well). For an AlAs quantum well, this form factor is given in~\cite{de2005effects, gold1987electronic}, with $C\approx 3$.

\subsection{Valley-dependent interaction terms}
In this subsection, we consider the contribution of inter-valley scattering terms that we neglected in treating the valleys as separate isospin flavours. In particular, let us consider the interactions within the full model, that takes into account the complete Brillouin zone with valleys around $X=(2\pi/a,0)$ and $Y=(0,2\pi/a)$ where $a_{\text{AlAs}}= 566 \rm{pm}$ the AlAs lattice constant. Here, we took into account that AlAs has a ``zinc-blend'' structure. Then, the inter-valley terms of the form
\begin{align}
v_{\ssm i-v}=\frac{1}{2L^2}\sum_{{\bf q}\neq 0} \bigg[ \tilde{v}({\bf q}) \sum_{\bf k} a^{\dagger}_{{\bf k} + {\bf q}} a_{\bf k} \sum_{\bf k'} a^{\dagger}_{{\bf k'-q}} a_{\bf k'}\bigg], \\
{\bf k} \in \text{valley $X$}, \ \ {\bf k+q} \in \text{valley $Y$}, \nonumber \\
{\bf k'} \in \text{valley $Y$}, \ \ {\bf k'-q} \in \text{valley $X$} \nonumber
\end{align}
are neglected when valleys are treated as isospin flavours.

The strength of this interaction term goes as $\tilde{v}({\bf q})$ with $q \sim \sqrt{2} \frac{2\pi}{a}$, where $a$ is the lattice constant ($q$ needs to connect between the $X$  and the $Y$ point of the Brillouin zone). Since for $q$ much larger than $1/w$ and $1/d$, $\tilde{v}({\bf q})\propto  3/(wq^2)$, this contribution is very small in comparison to intra-valley terms that involve a momentum transfer of $q \sim k_F$. Concretely, at a density $n=10^{11} \rm{cm}^{-2}$ (corresponding to $r_s \approx 15.4$, with $k_F \approx \sqrt{2\pi \cdot 10^{15}} \rm{m}^{-1}$), and using $w=20\,\rm{nm}$~\cite{hossain2021spontaneous}, the valley- (and spin-) dependent interactions are smaller by a factor of
\begin{align}
%\gamma \approx 
\frac{\tilde{v}(2\sqrt{2}\pi/a)}{\tilde{v}(k_F)} \approx \frac{3 k_F a^2}{8\pi^2 w}
%\frac{\tanh(d \sqrt{2} \frac{2\pi}{a})}{\tanh(d k_F)} 
\sim \mathcal{O}(10^{-4}).
\end{align}
This is small compared to the typical energy differences between states of different spin or valley polarization found in our calculations, which are of the order of $10^{-2}$ of the total energy (see Fig. 2 of the main text). We therefore conclude that the inter-valley scattering terms can be neglected.

This estimation is not significantly affected by the considered metal-gate screening, since
\begin{align}
 \frac{\tanh (d2\sqrt{2} \pi/a)}{\tanh (d k_F )}\approx 1,
\end{align}
for $d=100$ nm.

\subsection{Screening}
In modifying the gate distance $d$ in between $d=70$ and $d=300$ nm, we only found phase boundary shifts of similar order of magnitude as their error bars. However, we note that e.g. a single-gate screened potential or no screening at all could lead to more significant effects.

The evaluation and implementation of the dual-gate screened potential is further detailed in Sec.~\ref{dual-gateV}.

\subsection{Electron-phonon interaction}

We now discuss the effect of electron-phonon coupling, neglected throughout
most of this work. Coupling to the lattice favors the valley-polarized
state over the spin-polarized state, since the valley order parameter
couples linearly to an orthorombic lattice distortion. However, since
the 2DEG is embedded in a three-dimensional material, the energy gain
due to the distortion may expected to be small. Here, we show that
for parameters relevant to the experiment in Ref. \cite{hossain2021spontaneous}, the energy
gain due to the lattice distortion is negligible compared to the energy
differences we find in the purely electronic model (where only Coulomb
interactions are taken into account).

We add to the electronic Hamiltonian of Eq. (\ref{eq:H}) the following terms:
\begin{equation}
\Delta H = H_{el-ph}+H_{ph}.
\end{equation}
The electron-phonon coupling, $H_{el-ph}$, is written as
\begin{equation}
H_{el-ph}=\int d^{2}r\,\frac{1}{2}E_2 \left(n_{X}-n_{Y}\right)\left(\epsilon_{xx}-\epsilon_{yy}\right),
\end{equation}
where $n_{X},$ $n_{Y}$ are the two-dimensional densities of electrons
in the two valleys, $E_2$ is the shear tetragonal deformation potential, and $\epsilon_{ij}$ is the strain tensor.
The phonon Hamiltonian includes the elastic energy (we shall ignore
the phonon dynamics for simplicity):
\begin{equation}
H_{ph}=\int d^{3}r\:\frac{1}{2}\left[C_{11}\left(\epsilon_{xx}^{2}+\epsilon_{yy}^{2}\right)+2C_{12}\epsilon_{xx}\epsilon_{yy}\right],
\end{equation}
where $C_{11}$ and $C_{12}$ are elastic constants (we omitted the
third elastic constant characterizing a cubic system, $C_{44}$, since
it plays no role in our discussion). We start by assuming that the
2DEG has an effective thickness $w$, and that the strain is essentially
uniform across the thickness. (In reality, the thickness of the semiconductor
is much larger than the thickness of the 2DEG, as we shall discuss
below). The energy of the fully valley polarized state is then found
to be
\begin{equation}
E=\int d^{2}r\,\left\{\frac{1}{2}E_2\, n(\epsilon_{xx}-\epsilon_{yy})+\frac{w}{2}\left[C_{11}\left(\epsilon_{xx}^{2}+\epsilon_{yy}^{2}\right)+2C_{12}\epsilon_{xx}\epsilon_{yy}\right]\right\}.
\end{equation}
Here, $n$ is the two-dimensional electron density. 

Minimizing this expression over $\epsilon_{xx}$ and $\epsilon_{yy}$,
we find that the energy gain per unit area due to coupling to the
lattice is
\begin{equation}
E_{el-ph}=-\frac{E_2^{2}n^{2}}{4w\left(C_{11}-C_{12}\right)}.
\end{equation}
The energy gain per electron is $E_{el-ph}/n$. For AlAs, $C_{11}-C_{12}\approx 63\,{\rm GPa}$
\cite{adachi1985gaas} and $E_2 \approx 5.8\,\rm{eV}$ \cite{Charbonneau1991MeasurementElasticAlAs,Gunawan2006ValleySusceptibility}.
Taking $n=10^{11}\,{\rm cm}^{-2}$,
we obtain $E_{el-ph}/n\sim -10^{-3}\,{\rm meV}$ per electron.
This is significantly smaller than the typical energy differences we find in
the electron-only model between the valley and spin polarized states
(see Fig. 2 of the main text). We conclude that the effects of coupling
to the lattice can be ignored.

In fact, the thickness of the semiconductor in the experiment is much
larger than $w$. The lattice distortion is not uniform across the
thickness of the semiconductor. Moreover, depending on the thickness,
it may be favorable to form domains of opposite valley polarization,
such that the distortion decreases in the bulk as a function of distance
from the 2DEG. In any case, the fact that the semiconductor is thicker
than $w$ only decreases the energy gain due to the lattice distortion,
compared to the naive estimate given above.

\section{Optimization}

We use the variational principle
\begin{align}
E[\Lambda]:=\frac{\langle \Psi_{\Lambda} | \hat{H} | \Psi_{\Lambda} \rangle}{\langle \Psi_{\Lambda} | \Psi_{\Lambda} \rangle} \geq E_g,
\end{align}
where $E_g$ is the ground-state energy and $\Psi_{\Lambda}$ the trial wave-function with optimizable parameters ${\Lambda}$. By minimizing the expectation value of the Hamiltonian with respect to a given trial wave-function, one obtains a ground-state approximation.
Keeping the spin polarization fixed, the complexity in the evaluation of the above expectation value lies in a high-dimensional integral over all particle positions:
\begin{align}
\frac{\langle \Psi_{\Lambda} | \hat{H} | \Psi_{\Lambda} \rangle}{\langle \Psi_{\Lambda} | \Psi_{\Lambda} \rangle} = \frac{\int d{\bf r} d{\bf r' } \Psi_{\Lambda}^* ({\bf r}) \hat{H} \Psi_{\Lambda} ({\bf r'})}{\int d{\bf r} |\Psi_{\Lambda} ({\bf r})|^2 }= \frac{\int d{\bf r} |\Psi_{\Lambda} ({\bf r})|^2 H_L({\bf r}) }{\int d{\bf r} |\Psi_{\Lambda}({\bf r})|^2}.
\label{eq:Hexp}
\end{align}
Here, ${\bf r}=({\bf r}_1, ...{\bf r}_N)$, where ${\bf r}_i$ denotes the coordinates of particle $i$. 
We defined  the `local energy' $H_L({\bf r})=\Psi_{\Lambda}({\bf r})^{-1} \hat{H} \Psi_{\Lambda}({\bf r})$. The resulting integral can be efficiently estimated using Monte Carlo sampling: Instead of integrating over the complete Hilbert space, samples are drawn from the probability distribution $p\propto|\Psi_{\Lambda} ({\bf r})|^2$  using the Metropolis-Hastings algorithm. Then,
\begin{align}
\frac{\int d{\bf r} |\Psi_{\Lambda} ({\bf r})|^2 H_L({\bf r}) }{\int d{\bf r} |\Psi_{\Lambda}({\bf r})|^2} = \frac{1}{N_s} \sum_{{\bf r}_i \sim p}H_L({\bf r}_i)+\xi
\end{align}
and $\{ {\bf r}_i \}$ are the $N_s$ samples obtained via Metropolis Monte Carlo. The variance of the Gaussian-distributed statistical error $\xi$ with zero mean scales as $1/\sqrt{N_s}$, but vanishes when the Hamiltonian is evaluated and the trial wave-function corresponds to an eigenstate. Concretely, we can write this zero-variance property as \cite{casula2005new}
\begin{align}
\sigma^2=\sum_{{\bf r}_i \sim p}\bigg[H_L({\bf r}_i)-\bar{H}_L\bigg]^2 \geq 0, \\
\bar{H}_L=\frac{1}{N_s} \sum_{{\bf r}_i \sim p}H_L({\bf r}_i)
\end{align}
The equality holds when the wave-function corresponds to the ground-state (or an eigenstate), because then
\begin{align}
H_L({\bf r})=\Psi_{\Lambda}({\bf r})^{-1} \hat{H} \Psi_{\Lambda}({\bf r})=E_g, \ \ \ \bar{H}_L=E_g.
\end{align}
This zero-variance property allows for accurate estimation of the ground-state wave-function when an appropriate trial wave-function is used.

For minization of the variational energy $E[\Lambda]$, we use the stochastic reconfiguration technique introduced in
\cite{sorella1998green}, which can be understood as effective second order approximation to imaginary time evolution. 
The parameters $\Lambda$ of the trial wave-function are updated in every iteration as
\begin{align}
\Lambda \to \Lambda - \gamma S^{-1} F_{\Lambda},
\end{align}
where $\gamma$ is the learning rate and the force $F$ is given by
\begin{align}
F_k=2\bigg( \frac{\langle \partial_{\Lambda_k} \Psi_{\Lambda} | \hat{H}| \Psi_{\Lambda} \rangle}{\langle \Psi_{\Lambda}|\Psi_{\Lambda} \rangle}-E[\Lambda]\frac{\langle \partial_{\Lambda_k} \Psi_{\Lambda}| \Psi_{\Lambda} \rangle}{\langle \Psi_{\Lambda}|\Psi_{\Lambda} \rangle}\bigg).
\end{align}
The derivative with respect to the $k$-th variational parameter is denoted by $\partial_{\Lambda_k}$.

Second order effects are included by the covariance matrix \cite{park2020geometry}
\begin{align}
S_{k,k'}=\frac{\langle \partial_{\Lambda_k} \Psi_{\Lambda}| \partial_{\Lambda_{k'}}\Psi_{\Lambda} \rangle}{\langle \Psi_{\Lambda}|\Psi_{\Lambda} \rangle}-\frac{\langle \partial_{\Lambda_k} \Psi_{\Lambda}| \Psi_{\Lambda} \rangle  \langle \Psi_{\Lambda}| \partial_{\Lambda_{k'}}\Psi_{\Lambda} \rangle}{\langle \Psi_{\Lambda}|\Psi_{\Lambda} \rangle \langle \Psi_{\Lambda}|\Psi_{\Lambda} \rangle}
\end{align}
We employ the explicit regularization $S=S+\epsilon \mathbbm{1}$ in order to ensure invertibility. We use $\epsilon=10^{-3}$. At iteration $n_{\ssm it}$, we use the learning rate $\gamma={\rm max}(\gamma_0 \cdot 0.997^{n_{it}},\gamma_{\ssm min})$. The inital learning rate $\gamma_0$ is chosen (depending on density and system size) in the interval $[0.008,0.04]$ and $\gamma_{\ssm min} \in [0.0005,0.001]$.

\section{Trial wave-function}
The chosen parametrization of the trial wave-function, i.e. the concrete use of the parameters $\Lambda$, is of crucial importance to the accuracy of the results. The goal lies in representing the many-body correlated ground state wave-function as precisely as possible. At the same time, the trial wave function should also allow for efficient evaluation in order to keep the computational cost feasible.
In addition, antisymmetry has to be ensured for fermionic systems. An established route in ensuring antisymmetry lies in separating the wave-function into a product of a determinantal part, and a Jastrow factor
\begin{align}
\Psi_{\Lambda}({\bf r}_1, ...,{\bf r}_N)=\Psi_J({\bf r}_1, ...,{\bf r}_N)\cdot \Psi_D ({\bf r}_1, ...,{\bf r}_N), \label{eq:J_PsiD}
\end{align}
where the Jastrow factor $\Psi_J$ is chosen to be real,  positive and symmetric, such that the nodes and antisymmetry are fully determined by the determinantal part of the wave-function $\Psi_D$.
In particular, we employ a Slater-Jastrow-Backflow wave-function as elaborated in more detail below.

Before specifying more concretely the choice of Jastrow factor $J$ and determinantal part $\Psi_D$, we note here that in principle such a separation is not necessary. In particular, it has been shown that any totally antisymmetric wave-function can be represented as a single generalized determinant \cite{pfau2020ab, foulkesvariational}. The challenge lies then in finding the appropriate multi-electron orbitals. In \cite{pfau2020ab}, the authors have capitalized on deep neural networks as general function approximators \cite{bengio2017deep} to obtain these multi-electron orbitals. The resulting neural-network ansatz {\it Fermi Net} yields more accurate results for small atoms and molecules \cite{pfau2020ab} than any other wave-function to-date. However, a large computational cost is required to move to larger lattice sizes. 
For the isotropic electron gas, conventional trial wave-functions have been outperformed using a neural-network ansatz for systems with $27$ and $54$ electrons \cite{cassella2023discovering, wilson2022wave, li2022ab}.
The most scalable and accurate neural-network ansatz for periodic systems to-date constructed in \cite{pescia2023message} using message-passing neural networkds yields results for the $3D$ homogeneous electron gas up to $120$ electrons. While these consititute very promising results, many variational parameters and thus a comparably large computational cost is associated with neural-network quantum states for fermionic, continuous systems to-date. 
 
For the purpose of gaining physical insights about phases of multi-valley anisotropic systems, we are interested in probing the phase diagram as a function of $r_s$ and anisotropy on a dense grid. Thus, we require many simulations and use wave-functions of the product form \ref{eq:J_PsiD}, that are less accurate than the above mentioned wave-functions but only require little computational effort. 

We explain the form of our implemented trial wave-function and the use of the parameters $\Lambda$ in detail below.

\subsection{The Jastrow factor}
The Jastrow factor \cite{jastrow1955many} improves the many-body wave-function by effectively keeping electrons apart and creating a correlation hole.
We write the Jastrow prefactor of the wave-function as
\begin{align}
\Psi_J=e^{-J({\bf r}_1, ...{\bf r}_N)},
\end{align}
where the Jastrow factor $J$ is real and symmetric with respect to permutation of the particle positions $({\bf r}_1, ...{\bf r}_N)$.
While this non-negative bosonic prefactor \cite{holzmann2016theory} can in principle depend on all electron positions, in practice it is systematically constructed in a many-body expansion \cite{kim2018qmcpack}
\begin{align}
J=\frac{1}{2}\sum_{i,j} u_2 ({\bf r}_{i}, {\bf r}_{j}) +\frac{1}{6}\sum_{i,j, k} u_3 ({\bf r}_{i}, {\bf r}_{j }, {\bf r}_{k }) + ...,
\end{align}
with each $n$-body term being $u_n$ symmetric in the particle positions. Here, ${\bf }_{i}$ denotes the position of particle $i$ and the function $u_2$ ($u_3$) can depend on the isospins $\alpha_i, \alpha_j$ (and $\alpha_k$) of the particles $i$, $j$ (and $k$).

The two-body function $u_2$ has been approximated in early work as the random-phase-approximation (RPA) correlation function \cite{ceperley1978ground, kwon1993effects}. More expressive power is contained in the approximation of the two-body term as a (iso)spin-dependent liquid-like factor
\begin{align}
u_2({\bf r}_{i},{\bf r}_{j}) &=u_{\alpha_i, \alpha_j}(r_{ij}), \label{eq:u2}\\
r_{ij} &=|{\bf r}_{i}-{\bf r}_{i}|,
\end{align}
in combination with a polynomial expansion of $u_{\alpha_i, \alpha_i}(r_{ij})$ \cite{drummond2004jastrow, drummond2009phase} or a cubic Bspline interpolation as implemented in {\it qmcpack} \cite{kim2018qmcpack}:
\begin{align}
u_{\alpha_i \alpha_j}(r_{ij})=\sum \limits_{m=0}^M p_m B_3 \big( \frac{r_{ij}}{r_C /M}-m \big), \label{eq:ualphabeta}
\end{align}
where the cardinal cubic B-spline function $B_3(x)$ is centered at $x=-1$ and zero everywhere except on the interval $x\in [-3,1)$. The optimizable parameters are given by the $M$ control points $p_m$.
In order to comply with periodicity, the above parametrization is combined with cutting the Jastrow factor at the Wigner-Seitz radius $r_C$. Continuity of the wave-function and its first- and second derivatives is ensured by setting the last parameters $p_m$ to zero. The expressivity of the parametrization can be increased by increasing the number of control points $M$.

We note two issues with the above two-body Jastrow factor: First, a circular cutoff at the Wigner-Seitz radius limits the representability to short-range correlations, as the edges of the simulation cell are not included.
Second, the resulting correlation hole is isotropic. However, our Hamiltonian is anisotropic.

Both long-range correlations and anisotropic effects are typically encaptured by adding long-range periodic terms \cite{drummond2004jastrow}. In our case, we expect that the anisotropy in the kinetic energy of the Hamiltonian will have a relevant effect also on the short-range behaviour of the correlations. 
We thus propose the following two-body Jastrow factor to encapture both anisotropy and longer-ranged correlations
\begin{align}
u_2({\bf r}_{i},{\bf r}_{j})=u_{\alpha_i, \alpha_j}(r_{ij})+\nu_{\alpha_i, \alpha_j}({\bf r}_{i},{\bf r}_{j}).
\end{align}
Here, we separated the Jastrow factor into the short-ranged isotropic term $u$~(\ref{eq:ualphabeta}), and a second term $\nu$ allowing for anisotropy and correlations throughout the whole simulation cell. Any choice of $\nu$ that involves a cutoff and short-range anisotropy will either be restricted to very short-range correlations or lead to discontinuities as a consequence of periodic boundary conditions.
We thus resort to a different choice for the anisotropic term $\nu$. We combine anisotropy, cubic Bspline interpolation and a periodic ansatz presented in Ref.~\cite{whitehead2016jastrow}. In particular, we construct $\nu$ out of building blocks that already fulfill periodic boundary conditions \cite{whitehead2016jastrow}
\begin{align}
f_x({x}_{ij})=|{x}_{ij}|\big(1-2\frac{|{x}_{ij}|^3}{L_x^3}\big), \ \ 
f_y({y}_{ij})=|{y}_{ij}|\big(1-2\frac{|{y}_{ij}|^3}{L_y^3}\big),
\end{align}
with ${x}_{ij}=r^x_{i}-r^x_{j}$ being the x-component of the distance between particle $i$ and particle $j$. Respectively, $y_{ij}$ denotes the $y$-component. The length of the simulation cell in $x$ ($y$)-direction is given by $L_x$ ($L_y$), such that ${x}_{ij}\in (-L_x/2, L_x/2]$.  Here, we use periodic boundary conditions to define the distance within the Wigner-Seitz unit cell. The function $f_x$ is symmetric under $x\to -x$ and satisfies periodic boundary conditions at the edge of the simulation cell. More concretely
\begin{align}
f_x(L_x/2)=f_x(-L_x/2)\neq 0, \\
f'_x(L_x/2)=0, \\
f''_x(L_x/2)\neq 0. 
\end{align}
and vice versa for $f_y$. Thus, any function made out of $f_x$ and $f_y$ as building blocks automatically satisfies periodic boundary conditions.
In Ref.~\cite{whitehead2016jastrow} it was shown, that constructing a Jastrow factor out of a polynomial expansion of these building blocks can achieve lower energies for the homogenous electron gas than a combination of more conventional short-range and long-range Jastrow terms.

We make use of these building blocks $f_x$ and $f_y$ to design a Jastrow factor that can not only represent short-and long range correlations, but also anisotropy on all length scales. In addition, we find that numerical stability is improved using a combination with cubic Bspline interpolation instead of a polynomial expansion. We arrive at the Jastrow term 
\begin{align}\label{eq:parametrizationNugg}
\nu_{\alpha_i, \alpha_j}({\bf r}_{i},{\bf r}_{j}) &=A_1^{\alpha_i, \alpha_j}(g_1({\bf r}_{i},{\bf r}_{j}))+A_2^{\alpha_i, \alpha_j}(g_2({\bf r}_{i},{\bf r}_{j})), \\
g_1({\bf r}_{i},{\bf r}_{j})&=\sqrt{\lambda f_x^2 + (1-\lambda) f_y^2}, \label{eq:g1}\\
g_2({\bf r}_{i},{\bf r}_{j})&=\sqrt{(1-\lambda) f_x^2 + \lambda f_y^2}. \label{eq:g2}
\end{align}
Here, $A_1$ ($A_2$) corresponds to a cubic Bspline interpolation as in Eq.~(\ref{eq:ualphabeta}), defined on the image $g_1 \in [0,L_x/2]$ ($g_2 \in [0,L_y/2]$). No cutoff needs to be imposed here, and correlations in the whole simulation cell can be represented.
We used here that at small radius $r = \sqrt{x^2+y^2}$, $f_x(x)=|x|+\mathcal{O}(|x|^4)$, such that $(f^2_x(x)+f^2_y(y))^{1/2}=r+\mathcal{O}(|r|^4)$. The optimizable parameter $\lambda \in [0,1]$ therefore tunes the anisotropy: When $\lambda=0.5$, then the function $\nu$ is isotropic and $g_1({\bf r})=g_2({\bf r})=r+\mathcal{O}(|r|^4)$ for small $r$.

The used Jastrow factor components are isospin-dependent. With $\alpha=1, 2$ corresponding to spin-up and spin-down of valley $X$ and $\alpha=3, 4$ of valley $Y$, respectively, we captialize on the symmetries of the model by setting
\begin{align}
u_{11}=u_{22}=u_{33}&=u_{44},  \ \ \text{intra-valley, intra-spin} \label{eq:usym1}\\
u_{12}&=u_{34},  \ \ \text{intra-valley, inter-spin} \\
u_{13}=u_{14}=u_{23}&=u_{24}, \ \  \text{inter-valley.}
\end{align}
For the anisotropic $\nu$-term, we can make use of $90 \deg$ rotation between the two valleys and the parametrization of anisotropy detailed above, using $\lambda$. In particular, when denoting the parameter $\lambda$ explicitly in the function name by writing $\nu^{\lambda}$, we set
\begin{align}
\nu^{\lambda}_{11}=\nu_{22}^{\lambda}=\nu_{33}^{1-\lambda}=\nu_{44}^{1-\lambda}&, \ \  \text{intra-valley, intra-spin}\\
\nu_{12}^{\lambda}=\nu_{34}^{1-\lambda}&,  \ \ \text{intra-valley, inter-spin} \\
\nu_{13}^{\lambda}=\nu_{14}^{\lambda}=\nu_{23}^{\lambda}=\nu_{24}^{\lambda}&, \ \ \text{inter-valley}. \label{eq:vsymn}
\end{align}
Here, $\nu^{1-\lambda}$ corresponds to replacing the parameter $\lambda$ by $(1-\lambda)$ in the definition of $g_1$ and $g_2$ in Eqs.~(\ref{eq:g1})-(\ref{eq:g2}) and thereby effectively swapping $x$ and $y$.

\begin{figure*}[htbp]
  \centering
  \begin{minipage}[b]{0.45\textwidth}
    \centering
    \includegraphics[scale=1]{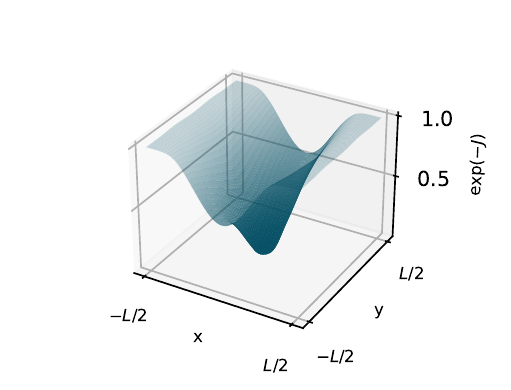}
    %\caption{Figure 1}
    %\label{fig:figure1}
  \end{minipage}
  \hfill
  \begin{minipage}[b]{0.45\textwidth}
    \centering
    \includegraphics[scale=1]{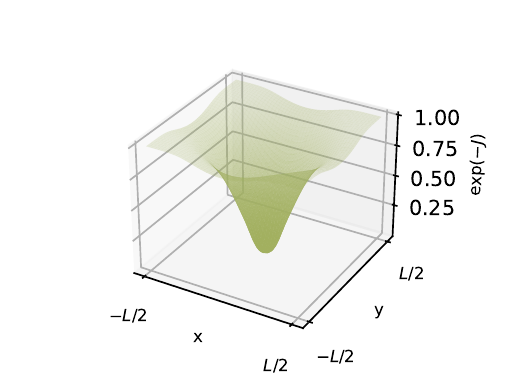}
    %\caption{Figure 2}
    %\label{fig:figure2}
  \end{minipage}
  \caption{Left: The Jastrow factor between two electrons of different spin in the {\em same valley}, for the VP state at $\eta=9$, $r_s\approx 21.7$ (simulation with $114$ electrons). The anisotropy of the valley is reflected in the Jastrow factor (correlation hole). Right: The Jastrow factor between two electrons of {\em different  valley}, for the SP state at $\eta=9$, $r_s\approx 21.7$ (simulation with $114$ electrons).}
  \label{fig:JastrowFactor}
\end{figure*}

Our Jastrow factor uses Bspline interpolation with $8$-$10$ control points (depending on density and anisotropy). Further increasing the number of segments did not result in lower energies. Figure~\ref{fig:JastrowFactor} shows the Jastrow factor between two electrons of different isospin optimized for the VP and SP state. The deviation from spherical symmetry can be observed in form of anisotropy, implying $C_2$-symmetry (intra-valley correlation, left hand side of Fig.~\ref{fig:JastrowFactor}) and $C_4$-symmetry (inter-valley correlation, right hand side of Fig.~\ref{fig:JastrowFactor}).

\subsection{The determinantal part}
The nodal structure is solely determined by the determinantal part of the wave-function.
Thus, the parametrization of the determinantal part should either already correspond to the correct nodal structure, or allow for flexibility such that optimization yields an accurate approximation thereof.

As a natural first step of {\it guessing} the nodal structure, we set the determinantal part to a single Slater determinant
\begin{align}\label{eq:DeterminatalPartDef}
\Psi_D({\bf r}_1,...{\bf r}_N) &=\prod \limits_{\alpha}\det \big( {\bf D}^{\alpha}\big), \nonumber \\
{\bf D}^{\alpha} &=\big( D_{ij}^{\alpha} \big)_{\{i,j\}}, \nonumber \\
D_{ij}^{\alpha} &= \phi_j^{\alpha}({\bf r}_i).
\end{align}
Here, we used a common trick to reduce computational cost and factorized the wave-function into a product of Slater-determinants ${\bf D}^{\alpha}$ per isospin sector $\alpha=(\sigma, \tau)$. Although the anti-symmetry of the wave-function is lost and we thus wrote down an unphysical state, this apparent issue dissolves when computing any observable diagonal in isospin basis. 

The single-particle orbitals $\phi^{\alpha}_{j}({\bf r}_i)$  of the involved Slater-determinants can then be computed by e.g. a mean-field approach. For closed-shell systems, single-determinant wave-functions often provide good approximations of the nodal surface. There exist several common approaches to improve upon such a wave-function. For small systems, a multideterminant wave-function can provide excellent results. However, the number of required determinants quickly becomes excessively high for large system sizes \cite{rios2006inhomogeneous}.
Instead, we here make use of Backflow transformations \cite{lee1981green, kwon1993effects, kwon1998effects}.

Formally justified in the context of Fermi liquid theory and the homogeneous electron gas \cite{kwon1993effects}, backflow transformation are introduced by evaluating the single-particle orbitals $\phi_j^{\alpha}({\bf r}_i)$ at a set of collective coordinates $({\bf x}_i, ...{\bf x}_j)$. More concretely, the determinal part of the wave-function with backflow transformations is then given by \cite{rios2006inhomogeneous}
\begin{align}
\Psi^{BT}_D({\bf r}_1,...{\bf r}_N) =\Psi_D({\bf x}_1,...{\bf x}_N),
\end{align}
where the new coordinates are given by
\begin{align}
{\bf x}_i={\bf r}_i+{\bf \xi}_i ({\bf r}_1,...{\bf r}_N).
\end{align}

The backflow displacement is typically parametrized using two-body interparticle distances \cite{lee1981green, kwon1993effects, schmidt1981structure, rios2006inhomogeneous}
\begin{align}
\xi_i^{\alpha_i \alpha_j} =\sum \limits_{j \neq i}^{N} \eta^{\alpha_i \alpha_j}_{ij} {\bf r}_{ij}, \\  {\bf r}_{ij} = {\bf r}_{i} -{\bf r}_{j} \label{eq:backflow}
\end{align}
where the number of electrons is given by $N$ and $\eta_{ij}^{\alpha_i \alpha_j}=\eta^{\alpha_i \alpha_j}(r_{ij})$ is a function that depends on the distance between two electrons, and their isospins $\alpha_i$, $\alpha_j$. We capitalize on the symmetries of the model with respect to the isospin in the same way as for the Jastrow factor, see Eqs.~(\ref{eq:usym1})-(\ref{eq:vsymn}). If the considered system includes nuclei, additional electron-nucleus and electron-electron-nucleus terms are typically added. Here, we consider all-electron systems such that the term Eq.~\ref{eq:backflow} is sufficient. 

We parametrize the two-electron function $\eta_{ij}$ using the very generic Bspline interpolation form implemented in {\it qmcpack} \cite{kim2018qmcpack}
\begin{align}
\eta^{\alpha_i \alpha_j}(r_{ij})=\sum \limits_{m=0}^M p_m B_3 \big( \frac{r_{ij}}{r_C /M}-m \big), \label{eq:backflowParam}
\end{align}
where the cardinal cubic B-spline function $B_3(x)$ is centered at $x=-1$ and defined on the interval $x\in [-3,1)$. The optimizable parameters are given by the $M$ control points $p_m$. Throughout or simulations, we use $8$ control points.
While the function $\eta$ is expected to decay as $r_{ij}^{-5/2}$ (for the 2D isotropic electron gas) \cite{kwon1993effects}, the above parametrization implies cutting the backflow function smoothly at the cutoff radius $r_C$. The advantages of this cutoff lie in computational efficiency and compatibility with periodic boundary conditions.
We however note here, that one could in principle employ a similar parametrization as we detail above for the Jastrow factor \cite{whitehead2016jastrow} in order to maintain computational efficiency while at the same time allowing to take advantage of the whole simulation cell. However, as the long-range effects of the backflow transformation are not as relevant as the correlations represented by the Jastrow factor, we keep the simple form \ref{eq:backflowParam}.

\subsubsection{Effective anisotropy}
Beyond the evaluation of the single-particle orbitals at generalized coordinates, the choice and filling of orbitals represents another important degree of freedom in the parametrization. Since we study fluid phases in a periodic simulation cell, we use plane-wave orbitals. The filling of these orbitals is, unlike the isotropic case, not protected by any symmetry. In particular, the effective shape of the Fermi surface may be renormalized by interactions. Since our model is based on a parabolic dispersion relation and the effective mass approximation, we parametrize this interaction-driven reshaping of the Fermi surface with a single parameter $\tilde{\eta}_{\ssm FS}$, that defines an ellipse in the same way as the bare anisotropy $\eta$. We do not optimize $\tilde{\eta}_{\ssm FS}$ using stochastic reconfiguration together with the other parameters as it determines the filling of the orbitals. Instead, we scan through a range of $\tilde{\eta}_{\ssm FS}$ and pick the solution with lowest energy.  In order to keep the computational cost low, the optimal effective anisotropy for each density and anisotropy is found using a Slater-Jastrow wave-function. Selected tests throughout the phase diagram were performed to confirm that this value is not changed (within the precision of the discrete $\tilde{\eta}_{\ssm FS}$-grid) when Backflow transformations are added.

In all our simulations, the effective anisotropy $\tilde{\eta}_{\ssm FS}$ is assumed to be the same in both valleys. In the case of the considered state with partial polarization (3 of 4 Fermi pockets filled, with the same density each), we note that no symmetry formally justifies this choice. 
However, the energy as a function of effective anisotropy has a plateau around the minimum for all states except the SVP (see Fig.~3 of the main part), where the change in energy is small in comparison to the energy difference to states with other isospin polarization. In addition, this plateau is in the same regime of $\tilde{\eta}_{\ssm FS}$'s for all states except the SVP. Thus, we do not expect a significant effect by allowing for different effective anisotropies in both valleys.

%{\it Trial moves.} When variational Monte Carlo is used for a spin system on a lattice, a local trial move can simply consist of flipping one spin. Here, the situation is more involved since one has to account for the continuous nature of the problem. In particular, particles can move around freely in the simulation cell. For the spatial part of the sample, we thus consider a local single-particle update drawn from the distribution
%\begin{align}
%T({\bf r}_0, ..., {\bf r}_i \to {\bf r}'_i, ...{\bf r}_{N})=\frac{1}{4\pi \tau} \exp \bigg( -\frac{({\bf r}'_i -{\bf r}_i)^2 }{4\tau}\bigg),
%\end{align}
%where the `timestep' $\tau$ is chosen such that the correlation length of the created samples in minimal - as a rule of thumb, this amounts to setting $\tau$ such that the Metropolis Monte Carlo acceptance probability is around $0.5$ \cite{casinomanual}.
%We use the above single-particle trial moves but note that more efficient sampling can be achieved by more involved sampling schemes, such as e.g. including a drift in the determination of a trial move \cite{kim2018qmcpack}.

\section{Evaluation of the Hamiltonian}
The ground-state energy is estimated by minimizing the variational energy, which requires evaluating the expectation of the Hamiltonian.
We detail below the evaluation of the considered Hamiltonian
\begin{align}
H=-\sum \limits_{i} \frac{1}{2m^{*}}\big(\eta^{\tau_i/2} \partial_{i,x}^2+ \eta^{-\tau_i/2} \partial_{i,y}^2 \big) + \sum \limits_{i<j} V(|{\bf r}_i- {\bf r}_j|),
\end{align}
consisting of an anisotropic valley-dependent kinetic energy and a dual-gate screened potential. The above Hamiltonian is defined in free space (in the thermodynamic limit). We further explain below the evaluation of the Hamiltonian with simulations performed in a finite simulation cell with periodic boundary conditions.

\subsection{Kinetic energy}
We perform simulations where each electron has a fixed isospin, since the Hamiltonian does not mix isospins. This approach has the advantage of reduced computational cost, as we do not need to keep track of a spinor part of the wave-function. However, the resulting wave-function is evidently unphysical. In particular, in the construction of the trial wave-function it is only possible to anti-symmetrize within the same isospin sector when keeping the isospin of each particle fixed. This apparent unphysical behaviour of the wave-function can be resolved when considering expectation values: For isospin-independent operators, the expectation value is the same as  the expectation value of the fully antisymmetrized version of this wave-function. However, the kinetic part of the Hamiltonian
\begin{align}
H_{\ssm kin}=-\sum \limits_{i} \frac{1}{2m^{*}}\big(\eta^{\tau_i/2} \partial_{i,x}^2+ \eta^{-\tau_i/2} \partial_{i,y}^2 \big)
\label{eq:Hkin}
\end{align}
is valley-dependent. We show below that this dependence does not pose an obstacly to the evaluation of the Hamiltonian with a trial wave function that is not antisymmetric with respect to exchange between isospin sectors.
Since all parts of the Hamiltonian are spin-independent we will for readability only focus on the valley degree of freedom. In particular, we define the valley eigenstates $|+1\rangle$ (electron in valley $Y$) and $|-1\rangle$ (electron in valley $X$). These are eigenstates to the operator $\hat{\tau}^z$:
\begin{align}
\hat{\tau}|\bar{\tau}\rangle=\bar{\tau}|\bar{\tau}\rangle, \ \ \bar{\tau}\in \{ +1, -1 \}.
\end{align}
Since we consider many-particle states with fixed isospin polarization, we introduce $N_{\ssm X}$ ($N_{\ssm Y}$) as the number of particles in eigenstate $|+1\rangle$ (eigenstate $|-1\rangle$).

Let us now re-write the kinetic part of the Hamiltonian~(\ref{eq:Hkin}) in more abstract terms. 
We can write it as a sum of the $x$- and $y$ part
\begin{align}
H_{\ssm kin}=H_{{\ssm kin},x}+H_{{\ssm kin}, y}
\end{align}
Both parts are of the same structure. In the following, we will just consider $H_{{\ssm kin},x}$ for simplicity, but the same arguments hold for the $y$-part.
Generally, $H_{{\ssm kin},x}$ consists of single-particle operators of the form
$\hat{O}_i f(\hat{\tau}_i^z)$, where $\hat{O}_i$ is an isospin-independent operator $\partial_{i,x}^2$ acting on particle $i$. $f(\hat{\tau}_i^z)$ is a function that involves the operator $\hat{\tau}_i^z$, corresponding to $\hat{\tau}^z$ acting on particle $i$. Here, $f(\hat{\tau}_i^z)=\eta^{ \hat{\tau}_i^z}$.
Then,
\begin{align}
H_{{\ssm kin},x}=\sum_n \hat{O}_n f(\hat{\tau}_n^z).
\label{eq:Hkin_f}
\end{align}

Now, we consider the spacial part of our trial wave-function $\phi (r_1, \tau_1, ...,r_n, \tau_n)$. By fixing the valley degree of freedom for each particle we do not have to keep track of the spinor part of the wave-function during the simulation. Including this part however explicitly results in the complete form of the trial wave-function
\begin{align}
\Psi_{\Lambda}(r_1, ... r_N)=\phi (r_1, \bar{\tau}_1, ...,r_n, \bar{\tau}_n) \zeta (\bar{\tau}_1, ...\bar{\tau}_N), \\
\zeta (\bar{\tau}_1, ... \bar{\tau}_N)=|\bar{\tau}_1\rangle ... |\bar{\tau}_N \rangle, \\
\bar{\tau}_1= ... =\bar{\tau}_{N_{\ssm X}}=+1, \ \ \bar{\tau}_{N_{\ssm X}+1}= ... =\bar{\tau}_{N_{\ssm X}+N_{\ssm Y}}=-1
\end{align}
We have used the notation $\bar{\tau}$ to distinguish the isospin {\em eigenstate} (label) from the position {\em variable}. Further, we will assume that $\Psi_{\Lambda}$ is antisymmetric with respect to exchange within the same isospin sector, but not between sectors.
Antisymmetry between sectors can be explicitly enforced:
\begin{align}
\Psi_{\Lambda}^{AS}(r_1, ... r_N)=\frac{1}{\mathcal{\sqrt{N}}}\sum_i (-1)^{{\rm sgn} (P_i)} \phi (r_{P_i(1)}, \bar{\tau}_{P_i(1)}, ...,r_{P_i(N)}, \bar{\tau}_{P_i(N)}) \zeta (\tau_{P_i(1)}, ...\tau_{P_i(N)}).
\label{eq:psiAS}
\end{align}
The sum goes over all $\mathcal{N}$ permutations $P_i$ that exchange particles between isospin sectors. 

We now want to show that
\begin{align}
\frac{\langle \Psi_{\Lambda} | H_{{\ssm kin},x} | \Psi_{\Lambda} \rangle}{\langle \Psi_{\Lambda}| \Psi_{\Lambda}\rangle}\overset{!}{=}\frac{\langle \Psi_{\Lambda}^{AS} | H_{{\ssm kin},x } | \Psi_{\Lambda}^{AS} \rangle}{\langle \Psi_{\Lambda}^{AS}| \Psi_{\Lambda}^{AS}\rangle}.
\end{align}
Using $\langle \Psi_{\Lambda}^{AS}| \Psi_{\Lambda}^{AS}\rangle=\langle \Psi_{\Lambda}| \Psi_{\Lambda}\rangle$, the above equation simplifies to
\begin{align}
\langle \Psi_{\Lambda} | H_{{\ssm kin},x} | \Psi_{\Lambda} \rangle \overset{!}{=}\langle \Psi_{\Lambda}^{AS} | H_{{\ssm kin},x } | \Psi_{\Lambda}^{AS} \rangle 
\label{eq:eqshow}
\end{align}
Re-writing the right hand side of the above equation using Eqs.~(\ref{eq:Hkin_f}) and (\ref{eq:psiAS}) yields
\begin{align}
\langle \Psi_{\Lambda}^{AS} | H_{{\ssm kin},x } | \Psi_{\Lambda}^{AS} \rangle =&\frac{1}{\mathcal{N}}\int dr_1 ... d r_N \sum_i (-1)^{{\rm sgn} (P_i)}\phi (r_{P_i (1)}, \bar{\tau}_{P_i(1)}, ...,r_{P_i (N)}, \bar{\tau}_{P_i (N)}) \zeta (\bar{\tau}_{P_i(N)}, ...\bar{\tau}_{P_i(N)} ) \nonumber \\ 
&\times\sum_n \hat{O}_n f(\hat{\tau}_n^z)  \sum_j (-1)^{{\rm sgn} (P_j)}\phi (r_{P_j (1)}, \bar{\tau}_{P_j(1)}, ...,r_{P_j (N)}, \bar{\tau}_{P_j (N)}) \zeta (\bar{\tau}_{P_j(N)}, ...\bar{\tau}_{P_j(N)})  =(*)
\end{align}

Using the orthogonality of the spinor basisfunctions $\zeta$ and the fact that they are $z$-eigenstates such that $\hat{\tau}_n^z\zeta(\bar{\tau}_1 ...\bar{\tau}_N)= \bar{\tau}_n\zeta(\bar{\tau}_1 ...\bar{\tau}_N)$, we obtain
\begin{align}
(*) =&\frac{1}{\mathcal{N}}\sum_i\int dr_1 ... d r_N \phi (r_{P_i (1)}, \bar{\tau}_{P_i(1)}, ...,r_{P_i (N)}, \bar{\tau}_{P_i (N)}) \zeta (\bar{\tau}_{P_i(N)}, ...\bar{\tau}_{P_i(N)} ) \nonumber \\ 
&\times\sum_n \hat{O}_n f({\bar{\tau}}_n)  \phi (r_{P_i (1)}, \bar{\tau}_{P_i(1)}, ...,r_{P_i (N)}, \bar{\tau}_{P_i (N)}) \zeta (\bar{\tau}_{P_i(N)}, ...\bar{\tau}_{P_i(N)}).
\end{align}
Using commutativity of addition and the integration order we arrive at
\begin{align}
(*) =&\frac{1}{\mathcal{N}}\sum_i\int dr_{P_i (1)} ... d r_{P_{i}(N)} \phi (r_{P_i (1)}, \bar{\tau}_{P_i(1)}, ...,r_{P_i (N)}, \bar{\tau}_{P_i (N)}) \zeta (\bar{\tau}_{P_i(N)}, ...\bar{\tau}_{P_i(N)} ) \nonumber \\ 
&\times \big( \hat{O}_{P_i(1)} f({\bar{\tau}}_{P_i(1)}) + ...+  \hat{O}_{P_i(N)} f({\bar{\tau}}_{P_i(N)})\big)  \phi (r_{P_i (1)}, \bar{\tau}_{P_i(1)}, ...,r_{P_i (N)}, \bar{\tau}_{P_i (N)}) \zeta (\bar{\tau}_{P_i(N)}, ...\bar{\tau}_{P_i(N)}) \\
=&\int dr_{1} ... d r_{N} \phi (r_{1}, \bar{\tau}_{1}, ...,r_{N}, \bar{\tau}_{N}) \zeta (\bar{\tau}_{N}, ...\bar{\tau}_{N} )  \big( \hat{O}_{1} f({\bar{\tau}}_{1}) + ...+  \hat{O}_{N} f({\bar{\tau}}_{N})\big)  \phi (r_{1}, \bar{\tau}_{1}, ...,r_{N}, \bar{\tau}_{N}) \zeta (\bar{\tau}_{N}, ...\bar{\tau}_{N})
\end{align}
where we have renamed integration variables in the last step and thus demonstrated the equality~(\ref{eq:eqshow}).
It directly follows that
\begin{align}
\frac{\langle \Psi_{\Lambda} | H_{{\ssm kin}} | \Psi_{\Lambda} \rangle}{\langle \Psi_{\Lambda}| \Psi_{\Lambda}\rangle}=\frac{\langle \Psi_{\Lambda}^{AS} | H_{\ssm kin } | \Psi_{\Lambda}^{AS} \rangle}{\langle \Psi_{\Lambda}^{AS}| \Psi_{\Lambda}^{AS}\rangle}.
\end{align}

%Maybe short: Evaluation as explained in Casino manual.

\subsection{Dual-gate screened interaction}
\label{dual-gateV}

In a simulation cell with periodic boundary conditions, the interaction between electrons has to be evaluated as a sum over all image charges:
\begin{align}
V &=\sum_{\bf L}  \sum_{i<j} V(|{\bf r}_i-{\bf r}_j + {\bf L}|)+ V_{\ssm Mad}+ V_{\ssm b}, \label{eq:Vsum} \\
V_{\ssm Mad} &=N\sum_{\bf L \neq 0} V(|{\bf L}|)
\end{align} 
The sum runs over all lattice vectors ${\bf L}$ of the simulation cell. The Madelung energy $V_{\ssm Mad}$ is a constant contribution that arises from interactions between particles and their own images. Furthermore, we consider all-electron systems. The contribution of the positive background is given by $V_{\ssm b}$, and corresponds to
\begin{align}
V_{\ssm b}=-\frac{1}{2}v({\bf q}=0)n^2 L^2,
\end{align}
where $n$ is the density, $L^2$ the area of the simulation cell in two dimensions and $v({{\bf q}=0})$ the $q=0$ component of the Fourier transform of the potential $V$.

While the above expression~(\ref{eq:Vsum}) is diagonal in real-space and can thus in principle be directly computed, the sum over all lattice vectors does not converge when long-ranged interactions such as the Coulomb potential are considered. The divergence cancels out with the (also diverging contribution) of the positive background, which cannot be translated into a real-space cutoff in the sum over lattice vectors.
This challenge is typically solved by breaking the interaction in a part that is short-ranged in real space and a part that is short-ranged in reciprocal space, for instance using Ewald summation \cite{sangster1976interionic, toukmaji1996ewald}.

For comparison with experimental observations however, implementation of a potential that is externally screened  may yield a more realistic comparison. Considering a two-dimensional electron system, if the potential is externally screened by two metal gates the sum over lattice vectors converges due to the exponential decay in real-space of the resulting potential \cite{throckmorton2012fermions}. Thus, it can directly be computed in real space. In addition, the positive background can be directly substracted since $v({\bf q}=0)$  is finite. Below, we derive the real-space form of the dual-gate screened potential and explain the concrete implementation in our simulations.

Concretely, we consider a two-dimensional electron system with two metal gates above and below, separated by distance $2d$. The gates induce screening of the Coulomb interaction within the electron system. The form of the obtained screened potential is well-known in Fourier space
%V(|{\bf r}_i-{\bf r}_j|)=\int {\rm d} {\bf q}\, {\rm e}^{{\rm i}{\bf qr}}v(|{\bf q}|), \\
\begin{align}
v({\bf q})=  \frac{e^2}{2 \epsilon_0 \epsilon} \frac{\tanh (d |{\bf q}|)}{|{\bf q}|},
\end{align}
Performing VMC calculations however requires access to the potential in real space. The Fourier transform
\begin{align}
V(r)=\frac{1}{(2\pi)^2}\int_{\mathbb{R}^2} {\rm d} {\bf q}\, {\rm e}^{{\rm i}{\bf qr}}v(|{\bf q}|)
\end{align}
is not analytically solvable. Instead, we directly compute the potential $V(r)$ in real space using the method of images. Concretely, we will repeat here the derivation given in \cite{throckmorton2012fermions}.

Two metal gates introduce infinite ``columns'' of image charges above and below the sample.
Concreteley, the resulting interaction potential between two electrons is given by \cite{throckmorton2012fermions, goodwin2020critical}
\begin{align}
V(r)=\frac{e^2}{4\pi \epsilon \epsilon_0} \sum_{n=-\infty}^{n=\infty}\frac{(-1)^n}{\sqrt{r^2+(2dn)^2}}.
\end{align}
Now, we proceed to compute the above series: We re-rewrite the sum by using the identity
\begin{align}
\frac{1}{r}=\frac{2}{\sqrt{\pi}}\int_0^{\infty} du e^{-r^2 u^2}.
\end{align}
Then, we obtain
\begin{align}
V(r)&=\frac{2}{\sqrt{\pi}}\frac{e^2}{4\pi \epsilon \epsilon_0 2d} \sum_{n=-\infty}^{n=\infty}\int_0^{\infty} du(-1)^n e^{-(r/(2d))^2u^2 -n^2u^2} \\
&=\frac{2}{\sqrt{\pi}}\frac{e^2}{4\pi \epsilon \epsilon_0 2d} \int_0^{\infty} du e^{-(r/(2d))^2u^2} \sum_{n=-\infty}^{n=\infty}(-1)^n e^{ -n^2u^2} \\
&= \frac{2}{\sqrt{\pi}}\frac{e^2}{4\pi \epsilon \epsilon_0 2d} \int_0^{\infty} du e^{-(r/(2d))^2u^2} \theta_4 (0, e^{-u^2}),
\end{align}
where we used the Jacobi theta function
\begin{align}
\theta_4(z,q)=\sum_{n=-\infty}^{\infty} (-1)^n q^{n^2} e^{2niz}.
\end{align}
Making use of the identity \cite{wolframurl} 
\begin{align}
\theta_4(z,q)=\frac{2\sqrt{\pi}}{\sqrt{-\ln q}}e^{(4z^2+\pi^2)/4\ln q} \sum_{k=0}^{\infty} e^{k(k+1)\pi^2/\ln q} \cosh \bigg( \frac{(2k+1)\pi z}{\ln q}\bigg)
\end{align}
we insert
\begin{align}
\theta_4(0,e^{-u^2})=\frac{2\sqrt{\pi}}{u} \sum_{k=0}^{\infty} e^{-(k+1/2)^2\pi^2/u^2} 
\end{align}
and obtain \cite{throckmorton2012fermions}
\begin{align}
V(r)&=4\frac{e^2}{4\pi \epsilon \epsilon_0 2d}\sum_{k=0}^{\infty}  \int_0^{\infty} du \frac{1}{u} e^{-(r/(2d))^2u^2} e^{-(k+1/2)^2\pi^2/u^2}  \\
&=4\frac{e^2}{4\pi \epsilon \epsilon_0 2d}\sum_{k=0}^{\infty} K_0 \bigg( (2k+1) \pi \frac{r}{2d}\bigg),
\label{eq:Vrscreened}
\end{align}
where we have evaluated the integral using modified Bessel functions of the second kind $K_n(x)$. The behaviour of the potential at large distances can be understood using the large-$x$ limit of $K_n(x)$:
\begin{align}
K_n(x) \approx \sqrt{\frac{\pi}{2x}}e^{-x}.
\label{eq:Blarge}
\end{align}
Thus, at large distance $r$
\begin{align}
V(r) \approx 4\frac{e^2}{4\pi \epsilon \epsilon_0 2d}\sum_{k=0}^{\infty}\sqrt{\frac{1}{(2k+1) r/d}}e^{-(2k+1)\pi r/(2d)} \sim \frac{1}{\sqrt{r}} e^{-\pi r/(2d)},
\end{align}
since the most dominant contribution for $r\gg d$ comes from the $k=0$ contribution and all other contributions are exponentially smaller. The decay of the dual-gate screened potential is thus approximately exponential in real space at large distances (with sub-leading pre-factor $1/\sqrt{r}$).
For numerical evaluation, we make use of the exact formula~(\ref{eq:Vrscreened}). In particular, due to the long-range behaviour of the modified Bessel functions~(\ref{eq:Blarge}), the sum can be truncated at $k\sim \mathcal{O}(d/r)$ and numerically computed. For very small $r$ when the evaluation becomes infeasible, the effect of screening is at the same time negligible and the bare Coulomb interaction can be used instead. 
In order to keep the computational cost low and avoid the evaluation of the sum~(\ref{eq:Vrscreened}) during runtime, we pre-evaluate the screened potential on a dense grid and load the stored values during the VMC simulations. In particular, we use an interval $r\in(\epsilon, l)$ where $\epsilon$ is very small and sufficiently $l$ large such that $V(l) \approx 0$. The potential is then evaluated at arbitrary $r$ by piecewise linear interpolation, if $r$ is in the pre-evaluated interval. For $r\geq l$, the potential is simply set to zero. For $r<\epsilon$, it is fitted to an $1/r$ behaviour chosen such that the overall potential is continuous (almost exactly corresponding to the bare Coulomb interaction). Here, we use $\epsilon \sim 2\times 10^{-3}a$, where $a$ is the lattice constant of AlAs. The length $l$ is an integer multiple of the simulation cell length, chosen such that $V(l)<10^{-10}$ meV.  

Further, for typical densities and particle numbers the screening length will be larger than one simulation cell. Thus, we evaluate the potential by summing over image cells. This sum given in Eq.~(\ref{eq:Vsum}) quickly converges due to the exponential decay of the interaction. Thus, we can truncate it after a finite number of terms. Throughout most of the simulations, we used the distance $d=100$ nm. For this value, the number of image cells required for evaluation of the potential ranges between $\mathcal{O}(1)$ to $\mathcal{O}(10)$ for the simulated densities and particle numbers. 

Due to the exponential decay of the potential in real space, the Madelung energy $V_{\ssm Mad}$ becomes negligible already at finite, sufficiently large system sizes. We thus directly set it to zero during the simulations such that there is no necessity to account for it in the finite-size scaling analysis.

We have varied the screening length in between $d=70$ and $d=300$ nm (for selected densities within the anisotropy $\eta=5.79$) and found that the effect on the estimated phase boundaries is of similar order as ambiguities in the phase boundary estimation resulting from phenomenological fitting functions. It is however possible that a comparably much enhanced screening or no screening at all could have a more enhanced effect on phase diagram.

We note however, that the effect of the varied screening length on the finite-size scaling - thus, on the finite-size contributions - is striking: The exponent (as explained in the next section) of the finite-size fit strongly varies with screening length.

\section{Finite-size scaling}

The goal of realistic numerical many-body approaches such as quantum Monte Carlo methods lies in producing reliable ground state properties. In the vast majority of use-cases such as the simulation of (here 2D) bulk systems this corresponds to systems close to the thermodynamic limit. However, calculations are necessarily performed with a finite number of electrons: 
For a given density $n$, the simulations are we use a finite number of electrons $N$ in a simulation cell of area $L^2=n/N$. Thus, the arising finite-size errors have to be accounted for.
Despite the existence of sophisticated methods for the correction of finite-size errors \cite{drummond2008finite, kwee2008finite, holzmann2016theory}, a largely successful practice in numerical studies of condensed matter consists of extrapolating finite-size energies to infinite system size, using an assumed relationship between energy and particle number.
Within QMC simulations, this relation typically takes into account scaling behaviour arising from finite-size corrections on the kinetic energy as well as resulting from a compression of the exchange-correlation hole into the simulation cell.
For a Fermi fluid within the isotropic 2D electron gas without external screening, careful consideration of the contributions to the scaling behaviour has resulted in the two-parameter form
\begin{align}
e_{\inf}=e(N)+a \Delta t_{\ssm HF}(N)+cN^{-\gamma},
\end{align}
introduced in \cite{ceperley1980ground}. Here, $e_{\inf}$ corresponds to the energy in the thermodynamic limit and is used as a fitting parameter together with $a$ and $c$. The finite-size energies are given by $e(N)$. The exponent $\gamma$ has been discussed in literature, the first result $\gamma=3/2$ \cite{ceperley1980ground} has been later improved to the more accurate $\gamma=5/4$ in \cite{drummond2008finite}.
$\Delta t_{\ssm HF}(N)$ corresponds to the difference in the infinite and finite-size kinetic energy within Hartree-Fock, and captures the finite-size effects that arise from discrete filling of the Fermi surface in metallic states. This effect can also directly be alleviated by considering offsets $(\Delta_x,\Delta_y)$ in the (2D) $k$-grid, which can be understood as twists in the simulation cell boundary conditions
\begin{align}
\Psi(x+nL,y+mL)=e^{{\ssm} i(n\theta_x+m\theta_y)}\Psi(x,y),
\end{align}
with $\theta_x=\Delta_x L$ and $\theta_y=\Delta_y L$. Pure periodic boundary conditions correspond to the choice $(\Delta_x,\Delta_y)=(0,0)$.
A smarter choice of the twist or averaging over random offsets (``twist averaging'' \cite{lin2001twist}) can be used to achieve $\Delta t_{\ssm HF}(N)=0$ directly.
Computational cost can be saved by considering particular twists. Concretely, we here implement the ``special twist'' method introduced in \cite{dagrada2016exact}: Simulations for a given state and electron number are performed with the same offset, which is chosen such that $\Delta t_{\ssm HF}(N)=0$. This condition does not uniquely define the special twist, leaving an ambiguity in the choice. However, it has been shown that different choices in the special twist lead to very similar results, as long as the condition $\Delta t_{\ssm HF}(N)=0$ is fulfilled \cite{dagrada2016exact}. We here choose offsets along the diagonal $(k_x,k_y)=(\alpha,\alpha)$ in order two treat both valleys on equal footing. Concretely, for a given density, number of particles, anisotropy and Fermi sea filling the diagonal $(\alpha,\alpha)$ is scanned and the condition $\Delta t_{\ssm HF}(N)=0$ checked for each value of $\alpha$. If a value of $\alpha$ can be found where the condition is fulfilled (for a chosen particle number), it is used for a simulation of the considered trial wave-function. We note that a special twist does not exist for all particle numbers and Fermi sea fillings, in particular when the effective anisotropy of the Fermi surface varies strongly from the bare anisotropy. However, in all our simulations we found a sufficient amount of special twists to perform finite-size extrapolation as further detailed below.  
% We note however, that a different choice (e.g. along the $x$-direction) leads to very similar results [].

\begin{figure}[tb]
\includegraphics[scale=1]{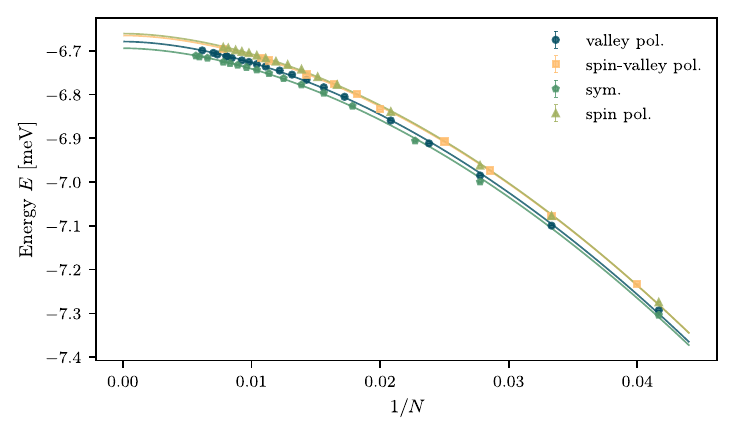}
\caption{Finite-size scaling at $\eta=3$, $r_s \approx 15.4$ for states with different isospin-polarization and a Slater-Jastrow-Backflow trial wave-function. In order to save computational cost, the simulated system sizes of the two lowest-lying states (up to $176$ electrons) are larger than for the states with higher energy. The statistical error bars are smaller than the size of the markers.}
\label{fig:scaling}
\end{figure}

We consider an anisotropic electron gas with dual-gate screened interaction and thus the exponent $\gamma=5/4$ derived for an unscreened potential does not apply here.
For very large lattice sizes, the exponential decay of the screened interactions is expected to result in all correlations decaying to zero within one simulation cell and thus a negligible finite-size error. For the considered lattice sizes however, the correlations surpass the size of the simulation cell and thus the regime of trivial behaviour in the finite-size error is not reached.
An accurate finite-size scaling behaviour requires careful consideration of the contributing corrections \cite{chiesa2006finite, drummond2008finite, holzmann2016theory}. We detail below that a rough estimation of the leading-order corrections is not sufficient for a well-behaved and reliable fit for the considered model, and instead extrapolate to the thermodynamic limit by including $\gamma$ as (anisotropy- and density-dependent) parameter that is fitted together with $e_{\inf}$ and $c$. The density-dependence of the exponent is effectively a result of finite-size-scaling contributions of different order with density-dependent prefactors. Despite the strongly phenomenological choice, the chosen fitting function yields a largely accurate fit when $r_s \lesssim 27$ (i.e. when the is system expected to be in a metallic phase, following the experimental predictions \cite{hossain2021spontaneous}) for the simulated system sizes, as shown exemplary in Fig.~\ref{fig:scaling} for $\eta=3$, $r_s \approx 15.4$. We note that we observe a more significant mismatch between simulated finite-size behavior and assumed phenomenological fit for the largest simulated $r_s \approx 30$, inducing a larger error in the finite-size scaling. We leave this observation to analyze in future studies. Further, we justify the phenomenological choice with our interest only in energy differences, which are less sensitive to the choice of finite-size extrapolation than absolute energies (with a sufficient amount of simulated particle numbers).

We detail below, the approximate estimation of contributions of the leading-order finite-size corrections and argue that they do not suffice for a reliable extrapolation.
As an instructing starting point for an estimate of the finite-size error on the potential $V_N$, we can write the electron-electron potential energy per particle for a simulation cell of area $L\times L$ containing $N$ electrons in Fourier space \cite{chiesa2006finite}
\begin{align}
V_N=\frac{1}{L^2}\sum \limits_{{\bf k}\neq 0} v({\bf k}) (\rho_{\bf k}\rho_{-\bf k}/N-1),
\label{eq:U_N}
\end{align}
where $\rho_{\bf k}:=\sum_j^N \exp (i {\bf k \cdot r}_j)$, and $v_k$ corresponds to the Fourier component of the dual-gate screened potential
\begin{align}
v({\bf k})=\frac{ e^2}{2 \epsilon_0 \epsilon}\frac{\tanh (d k)}{k},
\end{align}
with $k=|{\bf k}|$. The electron charge is given by $e$ and the metal-gate distance by $d$.
A finite-size error in Eq.~\ref{eq:U_N} is induced by the discrete $k$-mesh. As the system size increases, the mesh gets finer until the sum in (\ref{eq:U_N}) eventually converges to an inegral.

One can thus write the error on the potential energy per particle as the difference between integral and discrete sum
\begin{align}
\Delta V_N =\frac{1}{4\pi^2}\int v({\bf k}) (S({\bf k})-1) d{\bf k} - \frac{1}{L^2}\sum \limits_{\bf k \neq 0} v({\bf k}) (S_N({\bf k})-1),
\end{align}
where we used the static structure factor $S_N({\bf k})=\langle \rho_{\bf k} \rho_{-{\bf k}} \rangle/N$, and $S({\bf k})$ is the structure factor in the thermodynamic limit.

The leading order contribution is given by
\begin{align}
\Delta_1 =-\frac{1}{4\pi^2}\int v({\bf k}) d{\bf k}  + \frac{1}{L^2}\sum \limits_{\bf k \neq 0} v({\bf k}) .
\end{align}
 which is an integration error that arises due to the omission of the ${\bf k}=0$ area element from the sum. To leading order, the scaling of the error with number of particles can by estimated by considering this missing contribution, i.e. $\int_{\mathcal D} v_k d{\bf k}$, where $\mathcal{D}$ is a domain with area $4\pi^2/L^2$ \cite{chiesa2006finite}. However, $\Delta_1$ corresponds to the Madelung constant that we directly set to zero during our simulations, see Sec.~\ref{dual-gateV}. Thus, we do not need to account for the finite-size correction from the Madelung constant.

 The next-leading order correction comes from the discretization of $\int v({\bf k}) (S({\bf k})-1) d{\bf k}$ \cite{chiesa2006finite}. Approximating $S({\bf k})\approx S_N({\bf k})$ \footnote{This approximation is justified within the random-phase approximation. This becomes apparent when decomposing the potential into a long-range and short-range part. The long-range part that exhibits finite-size errors decays fast in reciprocal space. As a consequence, only the behaviour of $S({\bf k})$ at small $k$ is relevant. In the limit of $k\to 0$, the random phase approximation becomes exact}, one can use the same estimation as above for the integration error
\begin{align}
 \Delta_2 = \frac{1}{4\pi^2}\int v({\bf k}) S({\bf k})d{\bf k}  - \frac{1}{L^2}\sum \limits_{\bf k \neq 0} v_k S({\bf k})\propto \int_{\mathcal{D}}S({\bf k})v({\bf k}) d{\bf k}.
 \label{eq:Delta2}
 \end{align}
We note here that this estimation can not be applied for a quatitative computation of the finite-size correction, as it effectively implies neglecting a term of the same order of magnitude \cite{drummond2008finite}. We still apply the estimation, as we are only interested in the scaling behaviour. However we note that the scaling behaviour may also be affected by the missing contribution, as one would need to repeat the calculation in  \cite{drummond2008finite} for an externally screened potential.
 
Due to the validity of the random-phase approximation at small $k$, $S({\bf k})$ can be calculated analytically. For the isotropic 2D electron gas, $S({\bf k}) \propto k^{3/2}$ and $\Delta_2 \propto N^{-5/4}$. 
We here apply the random-phase approximation to compute the static structure factor $S({\bf k})$ in presence of a dual-gate screened interaction. We follow the derivation given in \cite{giuliani2005quantum} for the unscreened Coulomb potential and detail below only the differing steps using a dual-gate screened potential. Concretely, we make use of the relationship \cite{giuliani2005quantum}
\begin{align}
S({\bf k})=-\frac{\hbar}{\pi n}\int_0^{\infty} \mathcal{I}m \chi_{nn}( k,\omega) d\omega,
\label{eq:Sq}
\end{align}
where $\chi_{n n}(k,\omega)$ is the density-density response function. We use the approximation of the density-density response function obtained within the random-phase approximation
\begin{align}
\chi^{RPA}_{nn}(k,\omega)=\frac{\chi_0 (k,\omega)}{1-v_k \chi_0 (k,\omega)}.
\label{eq:chiRPA}
\end{align}
We introduced the notation $v_k :=v({\bf k})$ in order to simplify the consistency between vector and scalar objects.
Here, $\chi_0 (k, \omega)$ is the Lindhard function (density-density response function of independent electrons). 

In order to compute the imaginary part of the density-density response function, it is useful to search for its poles. We note here that the poles have also a direct physical meaning: The poles in the lower half of the complex frequency plane correspond to the frequencies of collective modes, giving rise to sharp peaks (resonances) in the spectral function. At long wavelength (small $k$), it is well-known that the spectrum of the density fluctuations within the random-phase approximation is dominanted by a collective excitation known as the {\it plasmon} \cite{giuliani2005quantum}.

We now turn to the calculation of the poles of $\chi^{RPA}_{nn}$.
They arise from zeros of the denominator of Eq.~(\ref{eq:chiRPA}), since the Lindhard function has no poles (only a branch cut along the real axis). Thus, the poles are given as solutions $\Omega_p$ of the equation
\begin{align}
1-v_k \chi_0 (k,\Omega_p)=0.
\label{eq:plasmonEq}
\end{align}
We use the small-$k$ expansion of the Lindhard function in two dimensions \cite{giuliani2005quantum} 
\begin{align}
\chi_0(k,\omega )\approx \frac{n k^2}{m \omega^2}[1+\frac{3}{4} \frac{k^2 v_F^2}{\omega^2}],
\end{align}
where $v_F$ is the Fermi velocity.
Then, the (real) solution to Eq.~(\ref{eq:plasmonEq}) is given by
\begin{align}
\Omega_p^2=\sqrt{\bigg(\frac{v_k n k^2}{2m}\bigg)^2+\frac{3}{4}k^4 v_F^2 \frac{v_k n}{m v_F}}+\frac{v_k n k^2}{2 m}.
\end{align}

We can then approximate the behaviour of the imaginary part of $\chi_{nn}^{RPA}(k,\omega)$ in the vicinity of the plasmon frequency $\Omega_p$ for $k\to 0$ as \cite{giuliani2005quantum}
\begin{align}
-\frac{1}{\pi}\mathcal{I}m \chi_{nn}^{RPA} (k,\omega) \approx \frac{\Omega_p(k)}{2 v_k} \delta (\omega-\Omega_p(k))
\end{align}
Inserting into Eq.~(\ref{eq:Sq}) yields
\begin{align}
S^{RPA}({\bf k})= \frac{\hbar}{n} \frac{\Omega_p (k)}{2 v_k}.
\end{align}
For the dual-gate screened interaction, the structure factor then approximately scales as $S({\bf k}) \propto k^{3/2} \tanh^{-1/2} (dk)$ in the small-$k$ limit.

We now want to determine the scaling of the correction $\Delta_2$ using the random-phase approximation of the structure factor. Inserting into Eq.~(\ref{eq:Delta2}), we obtain
\begin{align}
\Delta_2\propto  \int_0^{\frac{2\pi}{L}} dq \ 2\pi q S(q) v_q  \propto \int_0^{2\pi \sqrt{\frac{n}{N}}} q^{3/2}\tanh^{\frac{1}{2}}(dq). 
\end{align}
While this integral is not straightforward to evaluate, one can instead compute the integral numerically for a given density $n$ and gate-distance $d$ for a dense grid of values $1/N \in [0,1/N_{\ssm min}]$. We find that matching the obtained function with a fit $g(N)$ expanding in integer powers of $1/\sqrt{N}$ yields excellent agreement.
While we expect the anisotropy of the wave-function in principle also to enter in the finite-size correction, we did not include it in the above consideration due to the phenomenological observation that the scaling behaviour in our simulations is not very sensitive to the considered anisotropy.

Approximative scaling of the correction $\Delta_2$ is however not sufficient in order to obtain a reliable finite-size scaling. In particular, when employing the fit
\begin{align}
E_N=E_{\inf} - c  g(N),
\end{align}
the numerically obtained finite-size energies are not matched very well. In addition, we note that using an externally screened potential requires more careful consideration of the contributions to the finite-size error. More concretely, corrections from the discretized potential are much smaller than in the unscreened case: The potential is not divergent at small $k$, but finite. Thus, more subtle effects can become relevant such as further sub-leading corrections, dependence on the choice of trial wave-function \cite{holzmann2016theory} as well as higher-order corrections to the kinetic energy \cite{drummond2008finite}.

As we are further interested in energy differences, which are less sensitive to the choice of finite-size extrapolation than absolute energies (with a sufficient amount of simulated particle numbers), we choose to resort to the phenomenological fit $E_N=E_{\inf}-c N^{-\gamma}$ instead of more careful consideration of the finite-size corrections. The maximal particle numbers simulated are chosen with respect to the considered polarization, density and anisotropy. In particular, as shown in Fig.~\ref{fig:scaling}, the two states with lowest energy are simulated up to higher particle numbers (here $N=176$ and $N=162$ for the symmetric and VP state): Their energies are more relevant for the determination of the ground-state energy.
In principle, it is feasible to go simulate higher particle numbers $N>200$. Since we are constructing a phase diagram as a function of $r_s$ and $\eta$, and scan through a range of effective anisotropies at each point in the phase diagram, we perform our simulations with $N\leq 180$ to keep the overall computational cost manageable.

\section{Error bars}
In this section, we discuss the origin of the error bars on the phase boundaries (Fig. 1 in the main text) and on the extrapolated energies in the thermodynamic limit.

\begin{figure}[tb]
\includegraphics[scale=1]{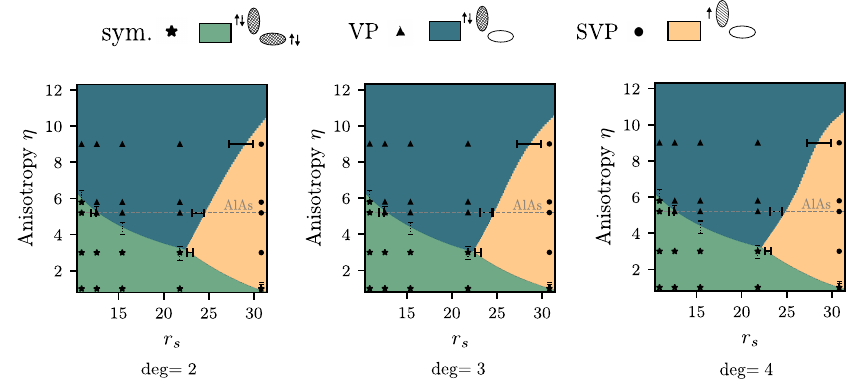}
\caption{Colored phase diagram for different degrees of interpolation polynomial in $\eta$-direction (vertical lines).}
\label{fig:phasediagram2D_deg}
\end{figure}

\paragraph{Error bar on extrapolated energies.} The precision of the variational energy, extrapolated to infinite particle number is given by error propagation of the statistical sampling error on the finite-size errors through the fit $e(N)=e_{\inf}-a \Delta t_{\ssm HF}(N)+cN^{-\gamma}$. Ambiguities in the fitting parameters contribute via the covariance matrix of the fit, but the phenomenological choice of scaling behaviour introduces an error that is here not accounted for.

\paragraph{Error bar on phase boundaries.}
Simulations are performed on a discrete grid in the $r_s$-and $\eta$-plane: We simulate the anisotropies $\eta \in \{1,3,5.2,5.79,9,12\}$ and the Wigner-seitz radii $r_s \in \{10.9, 12.6, 15.4, 21.8, 30.8\}$ (corresponding to the densities $n \in \{2.0, 1.5, 1.0, 0.5, 0.25\}\times 10^{11}$ cm$^{-2}$. The extrapolated energies (for each simulated isospin polarization) are fitted on horizontal and vertical lines in the plane, i.e. along $r_s$ and along $\eta$.
In particular, we use the parameterization of the correlation energy suggested by Rapisarda and Senatore \cite{rapisarda1996diffusion} to perfom a fit of the energies as a function of $r_s$. Albeit derived for an unscreened potential, we find the parametrization to yield an excellent fit to our energies. As a function of $\eta$, we employ a generic polynomial fit. Error propagation is used to obtain uncertainties in the fitting parameters of the vertical and horizontal fit. The parameters are then varied within this range, and the minimal and maximal position of the respective phase boundary define the error on its estimation. For a fit along a horizontal line, this error corresponds to an uncertainty in $r_s$. Vice versa, a fit on a vertical line yields phase boundary uncertainties in $\eta$. Since the order of the used polynomial on vertical lines is somewhat arbitrary, we use polynomials with order $2,3,4$, determine the uncertainty of fitting parameters for each choice of polynomial and determine the error on the phase boundary using the minimal and maximal estimated phase boundary of {\em all} performed polynomial fits. Figure~\ref{fig:phasediagram2D_deg} demonstrates the dependence of the phase boundaries on the choice of fitting polynomial: The error bars in all $3$ subfigures are determined as detailed above. The colors, however, are obtained by first performing a fit in $r_s$ and evaluating this fit on a dense grid of $500$ values in $r_s$. Then, the so-obtained energies are interpolated with a polynom of degree $2$ (left of Fig.~\ref{fig:phasediagram2D_deg}), degree $3$ (middle) and $4$ (right) in $\eta$-direction. Evaluating the polynomial fit again with $500$ anisotropies yields energies on a $500\times 500$ grid. These energies can be used to color the phase diagram, thereby providing a guide to the eyes.

An additional error arises from the discretization in the scanned effective Fermi surface anisotropies $\tilde{\eta}_{\ssm FS}$. Figure~4 in the main manuscript shows that this error is rather small for the VP, SP and symmetric state as these showcase approximately an energy plateau around the minimum. However, the SVP state shows a stronger dependence on the effective anisotropy. Thus, more accurate results can be obtained by using a finer $\tilde{\eta}_{\ssm FS}$-grid.

\section{Hartree-Fock}
We perform Hartree-Fock calculations directly in the thermodynamic limit. Hartree-Fock can be understood as a variational method, where the trial wave-function corresponds to a ground state as a noninteracting Hamiltonian. Here, we consider only states with fixed isospin-polarization. Then, the mean-field trial wave-function $\Psi_M$ is a Slater-determinant and the optimizable degree of freedom corresponds to the orbital filling, i.e. the shape of the effective Fermi surface.
As effects beyond a parabolic dispersion are neglected in our system, we parametrize the effective Fermi surface as an ellipse with anisotropy $\tilde{\eta}_{\ssm HF}$. Then, we use the variational principle
\begin{align}
\frac{\langle \Psi_M |H| \Psi_M \rangle}{\langle \Psi_M | \Psi_M \rangle}\geq E_g
\label{eq:HFvar}
\end{align}
to obtain a ground-state approximation.
We explain in the following the concrete steps of our algorithm.
First, we determine the above expectation value~(\ref{eq:HFvar}) given a state $\Psi_M$ with effective anisotropy $\tilde{\eta}_{\ssm HF}$ and a given isospin-polarization $(n_1, n_2, n_3, n_4)$. We denote the density in isospin flavour $\alpha$ as $n_{\alpha}$, and $\sum_{\alpha} n_{\alpha}=n$, where $n$ is the total density of the system. We adopt the notation, that the states $i=1$ and $i=2$ correspond to the spin-up, spin-down states of valley $\tau=1$, and $i=3$, $i=4$ to the valley $\tau=-1$. %In order to improve the readability of the following calculation, we will introduce $Tau :=\tau/2$.

\subsection{Kinetic energy}
The kinetic energy for a state with isospin polarization $(n_1, n_2, n_3, n_4)$ is given by
\begin{align}
E_{\ssm kin}=\sum_{{\bf k}, \alpha} \hbar^2\frac{ \eta^{\tau_{\alpha}/2} k_x^2+ \eta^{-\tau_{\alpha}/2} k_y^2}{2 m^{*}}n_{{\bf k},\alpha}.
\end{align}
Here, the isospin flavour is denoted with index $\alpha$. In addition, $n_{{\bf k},\alpha}=1$ if orbital ${\bf k}, \alpha$ is filled (and $0$ otherwise).
In the thermodynamic limit, we can replace the sum with an integral $\sum_{\bf {k}}\to L^2/(2\pi)^2 \int d {\bf k}$ and write
\begin{align}
E_{\ssm kin}=\frac{L^2}{(2\pi)^2}\sum_{\alpha}\int d {\bf k} \hbar^2\frac{ \eta^{\tau_{\alpha}/2} k_x^2+ \eta^{-\tau_{\alpha}/2} k_y^2}{2 m^{*}}\Theta((k^{\alpha}_F)^2-\tilde{\eta}_{\ssm HF}^{\tau_{\alpha}/2}k_x^2-\tilde{\eta}_{\ssm HF}^{-\tau_{\alpha}/2}k_y^2),
\end{align}
where the Fermi wave-vector for the isospin flavour $\alpha$ is given as $k_F^{\alpha}=\sqrt{4\pi n_{\alpha}}$.
With the substitution 
\begin{align}
\tilde{k_x}=\tilde{\eta}_{\ssm HF}^{\tau_{\alpha}/4}k_x, \ \
\tilde{k_y}=\tilde{\eta}_{\ssm HF}^{-\tau_{\alpha}/4}k_y, \ \ 
\end{align}
we obtain
\begin{align}
E_{\ssm kin}&=\frac{L^2}{(2\pi)^2}\sum_{\alpha}\int d {\bf k} \hbar^2\frac{ (\eta/\tilde{\eta}_{\ssm HF})^{\tau_{\alpha}/2} \tilde{k}_x^2+ (\eta/\tilde{\eta}_{\ssm HF})^{-\tau_{\alpha}/2} \tilde{k}_y^2}{2 m^{*}}\Theta((k^{\alpha}_F)^2-\tilde{k}_x^2-\tilde{k}_y^2) \\
&=\frac{L^2}{(2\pi)^2}\sum_{\alpha}\int_0^{k_F^{\alpha}} d\tilde{k} \tilde{k}^3 \int_0^{2\pi}  d\theta   \hbar^2\frac{ (\eta/\tilde{\eta}_{\ssm HF})^{\tau_{\alpha}/2} \cos(\theta)^2+ (\eta/\tilde{\eta}_{\ssm HF})^{-\tau_{\alpha}/2} \sin(\theta)^2}{2 m^{*}} \\
&=\frac{L^2}{(2\pi)^2}\pi \hbar^2\frac{ (\eta/\tilde{\eta}_{\ssm HF})^{\tau_{\alpha}/2} + (\eta/\tilde{\eta}_{\ssm HF})^{-\tau_{\alpha}/2} }{2 m^{*}}\sum_{\alpha}\frac{1}{4}(k^{\alpha}_F)^4 \\
&=L^2\pi\hbar^2\frac{ (\eta/\tilde{\eta}_{\ssm HF})^{\tau_{\alpha}/2} + (\eta/\tilde{\eta}_{\ssm HF})^{-\tau_{\alpha}/2} }{2 m^{*}}\sum_{\alpha}n_{\alpha}^2.
\end{align}
Thus,
\begin{align}
\frac{E_{\ssm kin}}{N}=\frac{\pi\hbar^2}{n}\frac{ (\eta/\tilde{\eta}_{\ssm HF})^{\tau_{\alpha}/2} + (\eta/\tilde{\eta}_{\ssm HF})^{-\tau_{\alpha}/2} }{2 m^{*}}\sum_{\alpha}n_{\alpha}^2.
\end{align}

\subsection{Exchange energy}
We adapt the derivation in \cite{giuliani2005quantum} of the exchange energy of an isotropic, two-dimensional electron gas for our case. The exchange energy is given by %Eq. (1.92) QTEL
\begin{align}
E_x=-\frac{1}{2L^2}\sum_{{\bf {q}}\neq 0} v_{\bf {q}} \sum_{{\bf k}\alpha}n_{{\bf k}+{\bf q} \alpha}n_{{\bf k} \alpha},
\end{align}
where $\alpha$ corresponds to the isospin flavour (spin $\sigma_{\alpha}$ and valley $\tau_{\alpha}$).

We proceed by computing the exchange energy.
Replacing the summation by integration (thermodynamic limit, $\sum_{{\bf k}}\to L^2/(2\pi)^2 \int d {\bf k}$) and using
\begin{align}
n_{{\bf k}\alpha}=\Theta((k^{\alpha}_F)^2-\tilde{\eta}_{\ssm HF}^{\tau_{\alpha}/2}k_x^2-\tilde{\eta}_{\ssm HF}^{-\tau_{\alpha}/2}k_y^2),
\end{align}
we obtain
\begin{align}
E_x= -\frac{L^2}{2(2\pi)^{4}}\sum_{\alpha}\int d {\bf q} \int d {\bf k} v_{{\bf q}} &\Theta((k^{\alpha}_F)^2-\tilde{\eta}_{\ssm HF}^{\tau_{\alpha}/2}(k_x+q_x)^2-\tilde{\eta}_{\ssm HF}^{-\tau_{\alpha}/2}(k_y+q_y)^2)    \nonumber\\
&\times\Theta((k^{\alpha}_F)^2-\tilde{\eta}_{\ssm HF}^{\tau_{\alpha}/2}k_x^2-\tilde{\eta}_{\ssm HF}^{-\tau_{\alpha}/2}k_y^2).
\end{align}
We can now make the following change of integration variables:
\begin{align}
\tilde{k}_x=\tilde{\eta}_{\ssm HF}^{\tau_{\alpha}/4}k_x, \ \
\tilde{k}_y=\tilde{\eta}_{\ssm HF}^{-\tau_{\alpha}/4}k_y, \ \ 
\tilde{q}_x=\tilde{\eta}_{\ssm HF}^{\tau_{\alpha}/4}q_x, \ \
\tilde{q}_y=\tilde{\eta}_{\ssm HF}^{-\tau_{\alpha}/4}q_y.
\end{align}
With $d\tilde{k}_x d\tilde{k}_y=dk_x dk_y$ (and likewise for $d {\bf q}$), we arrive at
\begin{align}
E_x= &-\frac{L^2}{2(2\pi)^{4}}\sum_{\alpha}\int d {\bf \tilde{q}} v_{{\bf q}}\int d {\bf \tilde{k}}  \Theta((k^{\alpha}_F)^2-(\tilde{k}_x+\tilde{q}_x)^2-(\tilde{k}_y+\tilde{q}_y)^2)    \nonumber\\
&\times\Theta((k^{\alpha}_F)^2-\tilde{k}_x^2-\tilde{k}_y^2),
\end{align}
where 
\begin{align}
v({\bf q})=v(\tilde{\eta}_{\ssm HF}^{-\tau_{\alpha}/4}\tilde{q}_x,\tilde{\eta}_{\ssm HF}^{\tau_{\alpha}/4}\tilde{q}_y).
\end{align}
We use that \cite{giuliani2005quantum} %(1.93) QTEL, careful sign mistake for 2D
\begin{align}
\frac{1}{N_{\alpha}}\sum_{{\bf k}} n_{{\bf k}+{\bf q} \alpha} n_{{\bf k}\alpha}=1-\frac{2}{\pi}\bigg[\arcsin (\frac{q}{2k^{\alpha}_F})+\frac{q}{2k^{\alpha}_F}\sqrt{1-\big(\frac{q}{2k^{\alpha}_F}\big)^2} \ \bigg]
\end{align}
for $q<2k^{\alpha}_F$ and $0$ otherwise. Here, $q=|{\bf q}|$.

\begin{figure}[tb]
\includegraphics[scale=1]{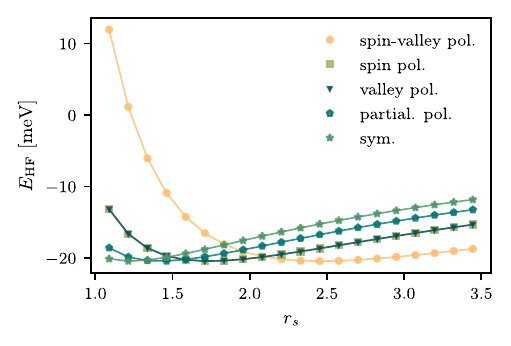}
\caption{Hartree-Fock energies in the thermodynamic limit for $\eta\approx 5.2$.}
\label{fig:HF}
\end{figure}

Then, we obtain
\begin{align}
E_x=-\frac{1}{2(2\pi)^{2}}\sum_{\alpha}\int_0^{2k^{\alpha}_F} d \tilde{q} \tilde{q} \int_0^{2\pi} d\theta v_{{\bf q}} N_{\alpha} \bigg(1-\frac{2}{\pi}\bigg[\arcsin (\frac{\tilde{q}}{2k^{\alpha}_F})+\frac{\tilde{q}}{2k^{\alpha}_F}\sqrt{1-\big(\frac{\tilde{q}}{2k^{\alpha}_F}\big)^2}\bigg] \bigg) .
\end{align}
Substituting $z=\tilde{q}/(2k^{\alpha}_F)$ and using that
\begin{align}
v({\bf q}) &=\frac{e^2}{2\epsilon_0\epsilon}\frac{\tanh(d\sqrt{q_x^2+ q_y^2})}{\sqrt{q_x^2+q_y^2}} \\
&=\frac{e^2}{2\epsilon_0\epsilon}\frac{\tanh(d\sqrt{\tilde{\eta}_{\ssm HF}^{-\tau_{\alpha}/2}\tilde{q}_x^2+\tilde{\eta}_{\ssm HF}^{\tau_{\alpha}/2}\tilde{q}_y^2})}{\sqrt{\tilde{\eta}_{\ssm HF}^{-\tau_{\alpha}/2}\tilde{q}_x^2+\tilde{\eta}_{\ssm HF}^{\tau_{\alpha}/2}\tilde{q}_y^2}} \\
&=\frac{e^2}{4 \epsilon_0\epsilon zk_F^{\alpha}}\frac{\tanh(2dzk_F^{\alpha}\sqrt{\tilde{\eta}_{\ssm HF}^{-\tau_{\alpha}/2}\cos(\theta)^2+\tilde{\eta}_{\ssm HF}^{\tau_{\alpha}/2}\sin(\theta)^2})}{\sqrt{\tilde{\eta}_{\ssm HF}^{-\tau_{\alpha}/2}\cos(\theta)^2+\tilde{\eta}_{\ssm HF}^{\tau_{\alpha}/2}\sin(\theta)^2}}.
\end{align}
Inserting into the exchange energy and dividing by the particle number, we obtain
\begin{align}
\frac{E_x}{N} = &-\sum_{\alpha}\frac{e^2 k_F^{\alpha}}{(2\pi)^{2}{2\epsilon_0\epsilon}}\frac{n_{\alpha}}{n}\int_0^{1} d z \int_0^{2\pi} d\theta  \frac{\tanh(2dzk_F^{\alpha}\sqrt{\tilde{\eta}_{\ssm HF}^{-\tau_{\alpha}/2}\cos(\theta)^2+\tilde{\eta}_{\ssm HF}^{\tau_{\alpha}/2}\sin(\theta)^2})}{\sqrt{\tilde{\eta}_{\ssm HF}^{-\tau_{\alpha}/2}\cos(\theta)^2+\tilde{\eta}_{\ssm HF}^{\tau_{\alpha}/2}\sin(\theta)^2}}  \\
 &\times \bigg(1-\frac{2}{\pi}\bigg[\arcsin (z)+z\sqrt{1-\big(z\big)^2}\bigg]\bigg) .
\end{align}
The above integral can be easily computed numerically for a given isospin polarization $(n_1, n_2, n_3, n_4)$.

\subsection{Hartree-Fock: Results}
We minimize $E/N=E_{\ssm kin}/N+E_x/N$ for the optimal effective anisotropy $\tilde{\eta}_{\ssm HF}$, for different isospin polarizations as a function of density $n$.

\begin{figure}[tb]
\includegraphics[scale=1]{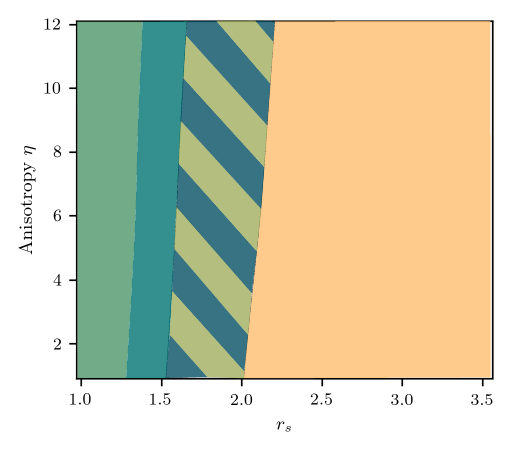}
\caption{Hartree-Fock phases in the thermodynamic limit as a function of $r_s$ and anisotropy. From left to right: Symmetric phase, partially polarized (3/4 polarized) phase, degeneracy between SP and VP, SVP phase.}
\label{fig:HFeta}
\end{figure}

Figure~\ref{fig:HF} shows the Hartree-Fock energies as a function of $r_s$ for states of different isospin polarization. We observe several attributes of the phase diagram that conflict with the experimental observations on AlAs: First, the valley-polarized, spin-unpolarized state (VP) and the spin-polarized, valley-unpolarized state (SP) are degenerate and thus mean-field cannot explain the experimental results on AlAs. 
The accidental degeneracy between VP and SP is a result of the form of the exchange energy: Only intra-valley terms contribute.
Second, the polarization transitions occur at much higher densities than observed in the experiment (VP in experiment at $r_s \approx 20$). Third, a phase with a partially polarized ground state with $3$ filled Fermi pockets exists within Hartree-Fock, which is also not observed in the experiment. The phase diagram as a function of anisotropy is given in Figure~\ref{fig:HFeta}. Here, we observe that the phase boundaries are rather insensitive to a change in anisotropy within Hartree-Fock, as opposed to VMC.

Above, we considered states with fixed isospin polarization. A  complete analysis that accounts for effects beyond polarized phases such as inter-valley coherence, requires including the possibility of isospin mixing. We will leave this analysis for future studies.

\bibliographystyle{unsrt}
\bibliography{papers.bib}

%TC:endignore